\documentclass[useAMS,usenatbib]{mn2e}
\usepackage{graphicx}
\usepackage{epsfig}

\title[Inner Disc Obscuration in GRS 1915+105]{Inner Disc Obscuration in GRS 1915+105 Based on Relativistic Slim Disc Model}
\author[K. Vierdayanti, A. S\c{a}dowski, S. Mineshige, and M. Bursa]{K. Vierdayanti$^{1}$\thanks{E-mail:kiki@as.itb.ac.id}, A. S\c{a}dowski$^{2}$, S. Mineshige$^{3}$, and M. Bursa$^{4}$\\
$^{1}$Department of Astronomy, FMIPA, Institut Teknologi Bandung, Ganesha 10, Bandung 40132, Indonesia\\
$^{2}$Harvard-Smithsonian Center for Astrophysics, 60 Garden St., Cambridge, MA 02138\\
$^{3}$Department of Astronomy, Kyoto University, Kitashirakawa Oiwake-cho, Kyoto 606-8502, Japan\\
$^{4}$Astronomical Institute, Academy of Sciences of the Czech Republic, Bocni II 1401, 141-31, Prague, Czech Republic}
\begin{document}

\date{Accepted 2013 00 00. Received 2013 00 00; in original form 2013 May 17}


\maketitle

\label{firstpage}

\begin{abstract}
We study the observational signatures of the relativistic slim disc of
$10\ {\rm M_{\odot}}$ black hole, in a wide range of mass accretion rate,	
$\dot{m}$, dimensionless spin parameter, $a_{\ast}$, and viewing angle, $i$.
In general, the innermost temperature, $T_{\rm in}$ increases with the increase
of $i$ for a fixed value of $\dot{m}$ and $a_{\ast}$, due to the
Doppler effect.
However, for $i>50^{\circ}$ and $\dot{m}>\dot{m}_{\rm turn}$, $T_{\rm in}$ starts to decrease with the increase
of $\dot{m}$. This is a result of self-obscuration --- the radiation from the
innermost hot part of the disc is blocked by the surrounding cooler
part.
The value of $\dot{m}_{\rm turn}$ and the corresponding luminosities depend on
$a_{\ast}$ and $i$.
Such obscuration effects cause an interesting behavior on the disc
luminosity ($L_{\rm disc}$) --
$T_{\rm in}$ plane for high inclinations.
In addition to the standard-disc branch which appears below $\dot{m}_{\rm turn}$
and which obeys $L_{\rm disc} \propto T_{\rm in}^{4}$ -relation,
another branch above $\dot{m}_{\rm turn}$, which is nearly horizontal,
may be observed
at luminosities close to the Eddington luminosity.
We show that these features are likely observed in a Galactic X-ray source, GRS 1915+105.
We support a high spin parameter ($a_{\ast} > 0.9$) for GRS 1915+105 since
otherwise the high value of $T_{\rm in}$ and small size of the emitting
region ($r_{\rm in} < 1 r_{\rm S}$) cannot be explained.
\end{abstract}


\section{Introduction}
Extensive studies of Galactic black hole binaries (BHBs) have shown that the
spectral transitions
can be related to the mass accretion rate
(e.g. Done, Gierli\'{n}ski \& Kubota 2007, Gilvanof 2010).
When the thermal component (disc blackbody) dominates the spectrum and
the luminosity is moderate
the disc is optically thick and geometrically thin.
In such a case, the accreting gas rotates with
Keplerian velocity and eventually reaches the innermost stable circular orbit
(ISCO) from where it rapidly falls towards the black hole. In this case,
ISCO may be identified with the
radiation inner edge
\citep{3}, since the radiation emitted
from inside the ISCO is negligible \citep{51}.
Note, however, that \citet{52} found that by using general relativistic magnetohydrodynamic simulation of accretion on to black holes, significant
emissivity can be expected inside ISCO radius.
In terms of
the Eddington luminosity, $L_{\rm Edd}$, such spectral state is commonly found
within luminosity range of $\sim$ 0.1 -- 0.3 $L_{\rm Edd}$ \footnote{See, however,
\citet{32}
in which they do not associate the spectral states with luminosity.}.
In this thermal state (sometimes referred to as the high/soft
state), the accretion flow can be described by the so-called standard disc model
\citep{35}. The relativistic version of this model was derived by
\citet{29}.

Using the standard disc model it is possible to derive the black hole
spin, disc luminosity and inclination from X-ray continuum fitting.
The black hole spin is generally estimated assuming that the disc
inner edge is at ISCO (see Shafee et al. 2006, and references therein).
Such an approach led to measurements of a wide range of black hole
spins, $a_{\ast}=0.0$--$1.0$
\citep{25}, which urges the need for theoretical models in which the
black hole spin is fully taken into account.
However, caution should be taken when one considers black hole systems
at the luminosity close to the Eddington
since the accretion disc is no longer properly
described by the standard disc theory.

As the mass accretion rate increases, the radiation pressure dominates
over the gas pressure and the cooling is provided by both vertical radiative
flux and horizontal advection
\citep{1}.
As a result, disc cannot maintain its thin geometry (becomes
thicker or slim) and the inner edge of the disc goes closer to the black
hole.
The disc becomes less radiatively efficient since the produced heat is not
totally emitted locally but some fraction of it is advected inward and emitted at
smaller radii or swallowed by the black hole. Therefore, radiation from inside
the ISCO may be not negligible and ISCO is no longer the radiation inner edge
(e.g. Watarai \& Mineshige 2003).
To study
such a flow,
\citet{1} proposed 'Slim Disc' model
which contains solutions corresponding to near-Eddington and moderately
super-Eddington accretion rates, $\dot{M} \ge \dot{M}_{\rm Edd}$, where
$\dot{M}_{\rm Edd}$ is the Eddington mass accretion rate defined as
\begin{equation}
\dot{M}_{\rm Edd}= \frac{L_{\rm Edd}}{c^2},
\end{equation}
where
\begin{equation}
L_{\rm Edd} = 1.25 \times 10^{38}\frac{M}{M_{\odot}} {\rm \ erg \  s}^{-1}
\end{equation}
is the Eddington luminosity.
That is,
\begin{equation}
\dot{M}_{\rm Edd}=1.39 \times 10^{17} \frac{M}{M_{\odot}} {\rm \ g \  s}^{-1}.
\end{equation}
For a non-rotating black hole, $\dot{M}_{\rm Edd}$ approximately
corresponds to the Eddington luminosity.
Hereafter, $\dot{m} \equiv \dot{M}/\dot{M}_{\rm Edd}$.

To date, slim disc model has been studied by many authors.
\citet{44}
discuss the slim disc model and its application to Galactic BHBs. Flatter
spectrum in the soft-X band as a result of flatter effective temperature
profile, $T_{\rm eff} \propto r^{-1/2}$, becomes the main feature that
distinguishes slim disc from that of the standard disc.
Watarai, Mizuno \& Mineshige (2001a)
study the application of slim disc model
for ultraluminous X-ray sources (ULXs). This is an interesting study since
the application of slim disc model for ULXs challenges the idea of intermediate
mass black hole (IMBH) for ULXs.
\citet{18}
investigates the application of slim disc model for narrow-line Seyfert 1
galaxy (NLS1) PG 1448+273 (see also Szuszkiewicz, Malkan \&
Abramowicz 1996; Wang \& Zhou 1999; Mineshige et al. 2000; Haba et al. 2008).
These studies, however, do not adopt a full-relativistic formalism, but
instead adopt the pseudo-Newtonian potential, $\psi = -GM/(r - r_{\rm
  S})$
\citep{31}, where $r_{\rm S}=2GM/c^2$ is the Schwarzschild radius.
In addition, they only consider a non-rotating (Schwarzschild) black hole case.
The relativistic effects are included in the spectrum calculation at best.

The observational evidence of the slim disc state elevates the recognition for
the slim disc model.
\citet{21} show that the slim disc state was observed in XTE J1550--564 data
during the high/soft state when the luminosity exceeds the critical luminosity.
In present days, slim disc model has been widely used to fit the spectra of many
luminous X-ray sources. The small inner radius ($<$ ISCO) and the flat temperature
profile obtained from the fitting are commonly claimed as the slim disc evidences
(e.g. Tsunoda et al. 2006; Okajima, Ebisawa \& Kawaguchi
2006). However, unlike the standard model, in which the radiation edge coincides
with the ISCO, the black hole spin estimate from the spectral fitting is not
straightforward.

\citet{33} revisited the study of slim disc around Kerr black holes.
They focused on the parameters which are applicable to stellar mass BHBs.
More importantly, fully relativistic slim disc model provides a tool
to study the signatures of black hole spin in the spectra.
In \citet{33}, however, the effect of wind-mass loss is not taken into account.
At high mass accretion rate, wind-mass loss becomes important, due to significant radiation pressure (Takeuchi, Mineshige \& Ohsuga 2009).
However, as Takeuchi et al. (2009) have shown, the temperature and scale height radial profiles of the slim disc are hardly affected by non-constant profile of mass accretion rate.
In the present study, we try to bridge the gap between theory and
observations by fitting the relativistic slim disc spectra with the extended
disc blackbody (extended DBB) model (also known as $p$-free model;
Mineshige et al. 1994).

Unlike mass,
black hole spin is much more difficult to measure since it only affects the
region very close to the event horizon and thus much less observable from
a large distance. Nevertheless, there have been several attempts to measure
the black hole spin by using the X-ray spectral fitting (see McClintock et al.
2010, and references there in). The disc spectra in
these studies, however, are assumed to be in the standard thermal
state which
can not be suitable for high mass accretion rate systems.
For this reason, \citet{24}, carefully selected data with luminosity below 30\% of the
Eddington luminosity.
To overcome this limitation,
fitting with
relativistic slim disc model will be important.

To understand the observed properties of black hole accretion
flows shining at around the Eddington luminosity, we fit the
synthetic spectra of high mass accretion rate discs based on the fully relativistic
slim disc model with the extended DBB fitting model.
Assuming that the relativistic slim disc model is a better representation of
the realistic situations, we aim
(1) to give a new methodology to evaluate black hole spins from
the X-ray spectral fitting of the luminous objects whose
luminosity exceed $\sim 0.3 L_{\rm Edd}$, and (2) to give reasonable
interpretation to the peculiar behavior of GRS 1915+105 in the
$L_{\rm disc}$ -- $T_{\rm in }$ diagram (Vierdayanti, Mineshige \&
Ueda 2010).

The plan of this paper is as follows:
We will review the relativistic slim disc model
\citep{33}
and the spectrum calculations in section 2.
In section 3, we investigate the
effects of the mass accretion rate, black hole spin
and viewing angle on the extended DBB model parameters.
We present the comparison of our study with the
observations of GRS 1915+105 in section 4.
In section 5, we discuss some remaining issues from our
study, such as the interpretation of $T_{\rm in}$, the
disc geometry and the black hole spin. We conclude
the paper in section 6.

\begin{figure*}
 \includegraphics[angle=270,scale=0.62]{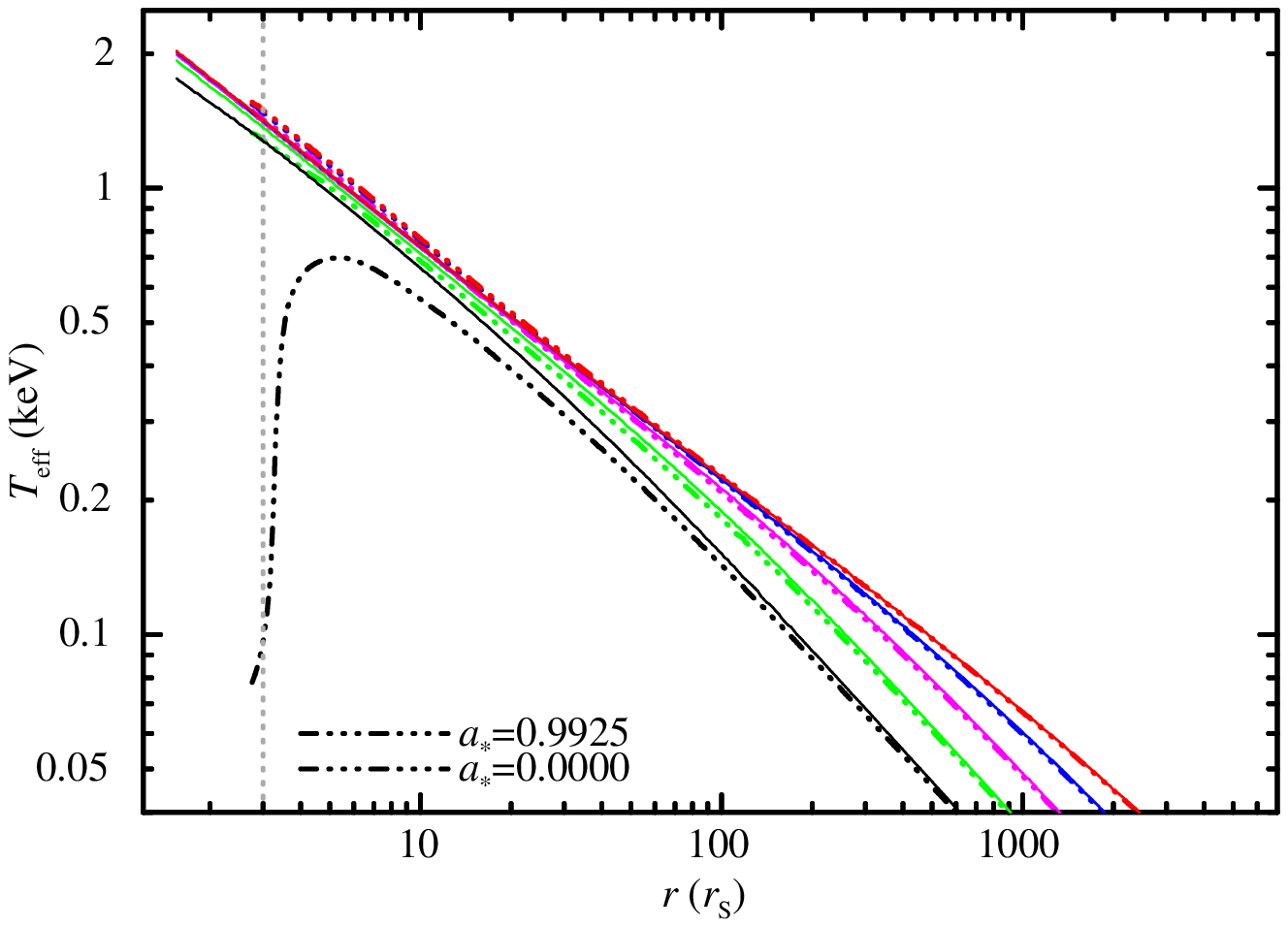}
 \includegraphics[angle=270,scale=0.62]{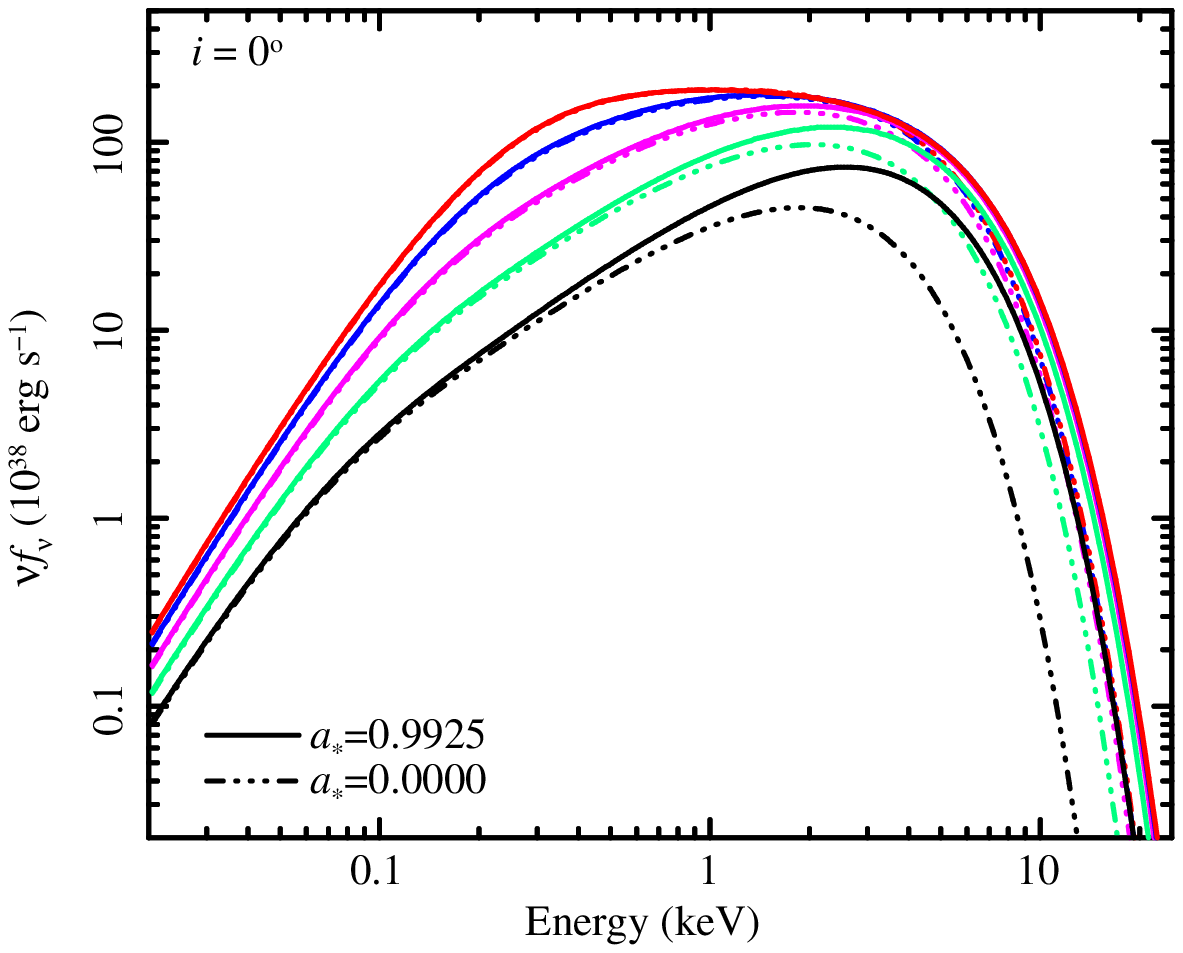}
  \caption{Temperature profile (left) and the spectra (right) of relativistic slim disc for various mass accretion rate.
Here, we only show for $\dot{m}=10$, 32, 100, 320, \& 1000 in black, green,
magenta, blue and red, respectively, for clarity. Solid lines: $a_{\ast}=0.9925$,
dash-dot-dot-dot lines:$a_{\ast}=0.0000$. The grey vertical dotted line in the left panel marks $r=3r_{\rm S}$.
}
\label{Figure:1}
\end{figure*}
\begin{figure*}
 \includegraphics[angle=270,scale=0.65]{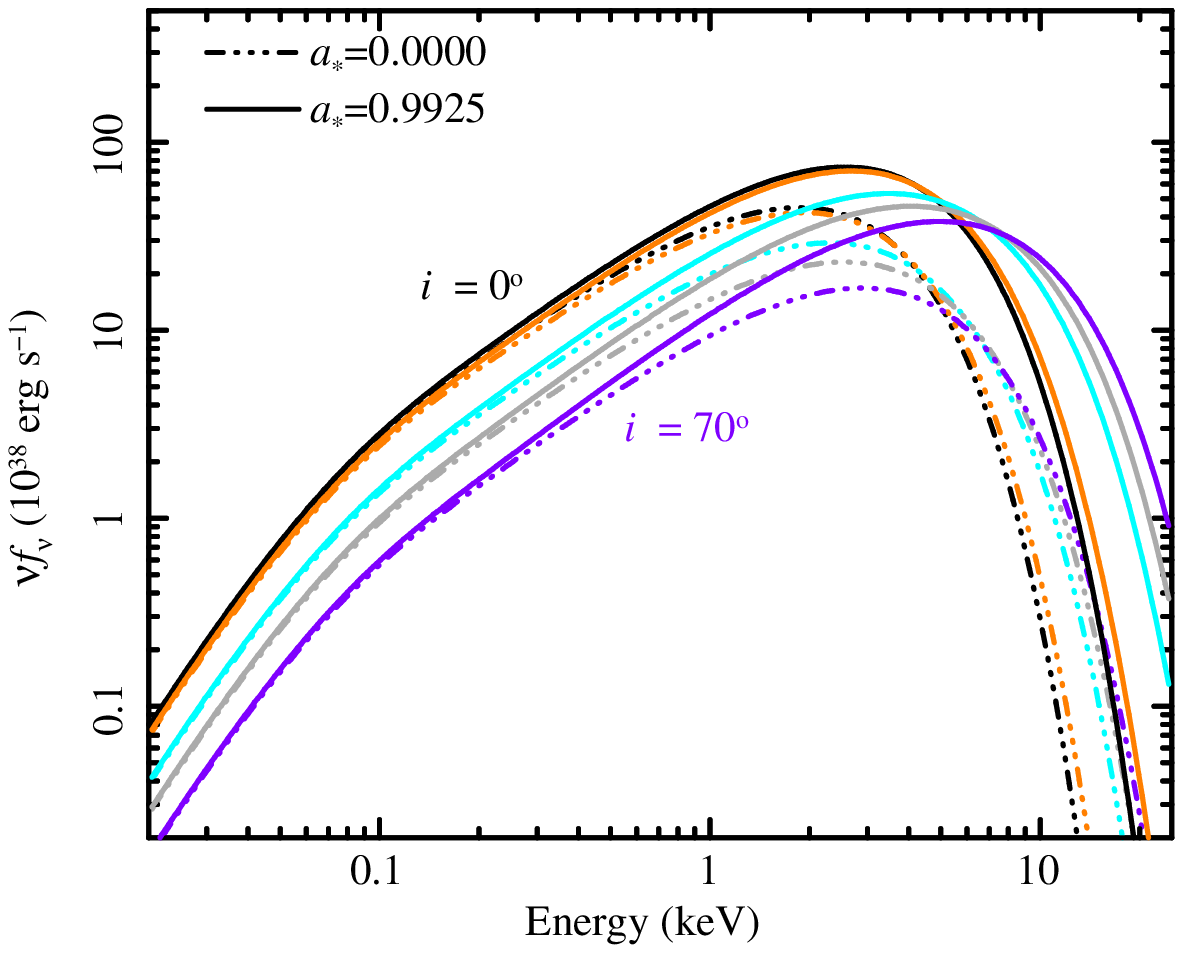}
 \includegraphics[angle=270,scale=0.65]{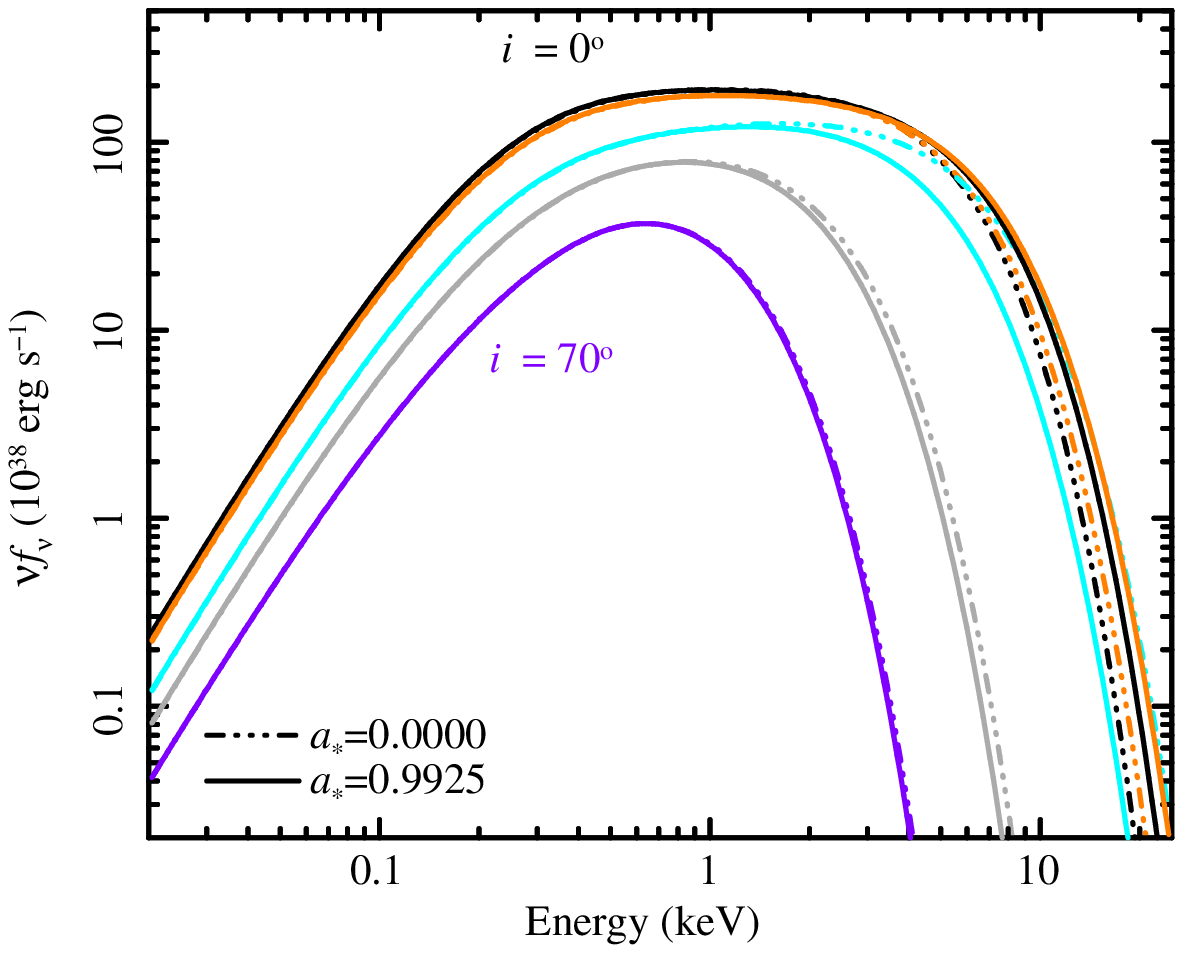}
  \caption{Spectra of relativistic slim disc for various viewing angles:
$i=0$, 20, 50, 60 \& $70^{\circ}$ in black, orange, cyan, light gray and purple,
respectively, for $\dot{m}=10$ (left) and $\dot{m}=1000$ (right).
}
\label{Figure:2}
\end{figure*}
\section[]{The Relativistic Slim Disc Properties and the Spectral Features}
Slim disc model was introduced in the late 1980s by the
Warsaw and the Kyoto groups (Abramowicz et al. 1988; Abramowicz,
Kato \& Matsumoto 1989; see Kato, Fukue \& Mineshige 2008 chapter 10 for a review).
Slim discs are astrophysically important since they can be applied to
accretion flows with $1 \le \dot{M}/
\dot{M}_{\rm Edd} \le 100$
\citep{2}.
This range of accretion rate has been observed in some Galactic and
extra-Galactic X-ray sources.

In this work we use the slim disc solutions around Kerr black
hole obtained by
\citet{33}, hereafter S11.
We will only review the main results of this work while the details can be
found in the paper.

\subsection{Relativistic slim disc properties}
When the luminosity approaches or moderately exceeds the Eddington
luminosity (equation (1)), the generated energy can no longer be totally emitted
locally due to photon trapping.
Some fraction of the generated photons are trapped by the gas and advected inward due to
high (scattering) optical depth.
The slim disc solutions deviate from that of the standard disc when advection becomes important. Fig. 4.11 of S11 shows the increase of heat advected with the increase of mass accretion rate. Some part of the advected
heat will go too close to the black hole and cannot escape. Some other part will be emitted in the smaller radii so that it will amplify the local emission by the viscous
process.
As shown in fig. 4.2 of S11, the peak in the radial profiles of the flux for
the higher mass accretion rates is shifted to the smaller radii.
In the case of highly spinning black hole, the shift is
particularly less visible due to the common shift caused by the decreasing
radius of the ISCO with the increasing spin parameter.

The $H/r$ ratio, where $H$ is the disc scale height and $r$ is the disc radii,
and the central temperature profiles are consistent with previous studies
(see fig. 4.5 of S11).
As the mass accretion rate increases, the radiation pressure
dominated region is formed in the inner region of the disc which results
in an increase of the $H/r$ ratio.
The increase of the $H/r$ ratio produces significant effects on the spectra
which will be described in the next subsection.
The temperature of the gas pressure
dominated region increases more rapidly with decreasing radius than that in
the radiation pressure dominated region. That is, the radial profiles of the
temperature at the inner region become flatter as mass accretion rate increases.

For moderately high accretion rates, i.e. the case of optically thick
and geometrically thin disc, the energy conversion efficiency $\eta=L/
\dot {M} c^2$ is constant (does not depend on
the accretion rate), since the luminosity increases proportionally with
$\dot{m}$. When the accretion rate increases, the inner edge of the disc
leaves the ISCO and moves inward.
In addition, photon trapping effect becomes important that suppresses the
photons from being emitted from the surface of the disc.
Therefore, the luminosity is not proportional to $\dot{m}$. Instead,
it grows slower than $\dot{m}$ (S11 and references therein).

\subsection{The observed spectra of relativistic slim disc model}
In the previous studies (e.g. Watarai et al. 2000, 2001a, 2001b, hereafter
W00--01), where the pseudo-Newtonian potential was adopted and the Schwarzschild
(non-rotating) black hole was assumed, the slim disc model was found to exhibit
several important signatures:
the spectra are multicolor blackbody characterized by
a high maximum temperature ($T_{\rm in}\sim$ a few keV), a small emitting region
($r_{\rm in} < 3r_{\rm S}$), and flatter spectra in the soft X-ray band due to a
flatter effective temperature profile.

The slim disc model used in our study is constructed in the Kerr space-time
metric and the spectra include all special and general
relativistic effects
\citep{6}.
The ray-tracing method is used to calculate the emission from the proper effective photosphere to an observer at infinity.
The deviation of the observed spectra from a pure multi-color blackbody, due to the change in the disc opacity, is taken care by a spectral hardening factor, $\kappa$, which is the ratio between the color temperature and effective temperature.

Fig. 1 shows the temperature profile (left panel) and spectra (right panel) of relativistic slim disc in 5 different mass accretion rates,
for the case of $a_{\ast}=0.0000$ (dash-triple-dot lines) and
$a_{\ast}=0.9925$ (solid lines).
Colors represent the mass accretion rate
and will be used again in Fig. 5 \& 7. Black, green, magenta, blue, and red denote $\dot{m}=10$, 32, 100, 320,
and 1000, respectively.
In agreement with previous studies, as mass accretion rate increases, temperature profile becomes flatter.
Flatter temperature profile produces flatter spectrum in the soft energy band.
The total flux increases as $\dot{m}$ increases for both non-rotating and
rapidly rotating black hole case.
The spectra for $a_{\ast}=0.9925$ differ significantly from those of
$a_{\ast}=0.0000$ at X-ray energy range, especially at $> 1$ keV, when $\dot{m}=10$.
The difference becomes less noticeable as $\dot{m}$ increases.
In general, the spectrum becomes harder as the spin parameter increases.

Fig. 2 shows the spectra of various viewing angles $i=0$, 20, 50, 60 and
$70^{\circ}$ for $\dot{m}=10$ (left) and for $\dot{m}=1000$ (right).
Similar to Fig. 1, the dash-triple-dot and solid lines correspond to
$a_{\ast}=0.0000$ and $a_{\ast}=0.9925$, respectively.
Colors represent different $i$ and will also be used in
Fig. 3, 4 \& 6. Black, orange, cyan, light gray, and purple represent
$i=0$, 20, 50, 60, \& $70^{\circ}$, respectively.

As shown in Fig. 2, for $\dot{m}=10$ (left panel), the spectrum becomes
harder and the total flux decreases, as $i$ increases.
In the case of $\dot{m}=1000$ (right panel), for $i>0^{\circ}$, the spectrum
becomes harder and the total flux decreases, similar to the $\dot{m}=10$
case. However,
at a certain point ($i \sim 50^{\circ}$), the spectrum becomes softer as $i$
increases further, in opposite to the face-on case. The total flux,
on the other hand, decreases significantly.
\begin{figure*}
 \includegraphics[angle=270,scale=0.8]{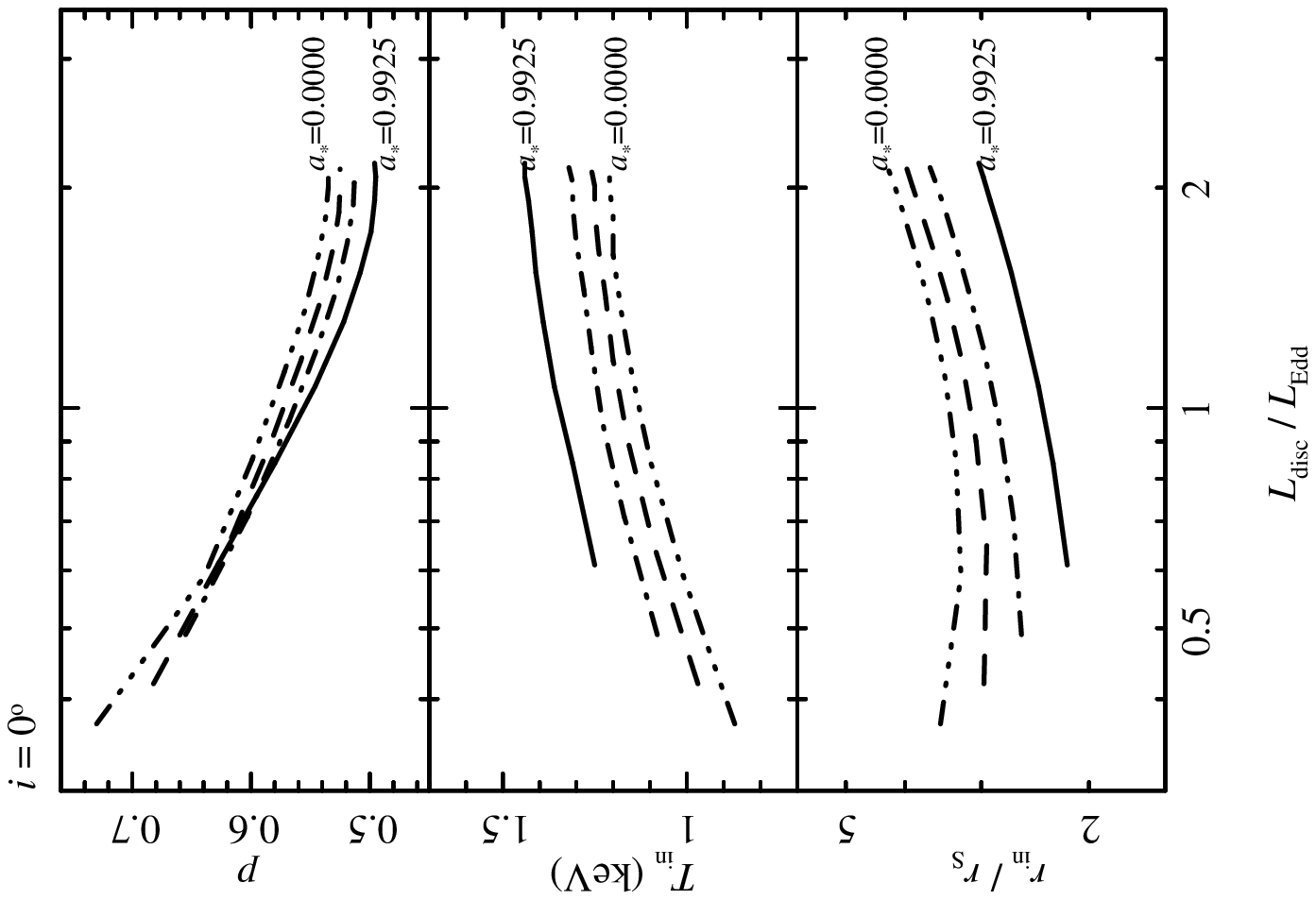}
 \includegraphics[angle=270,scale=0.8]{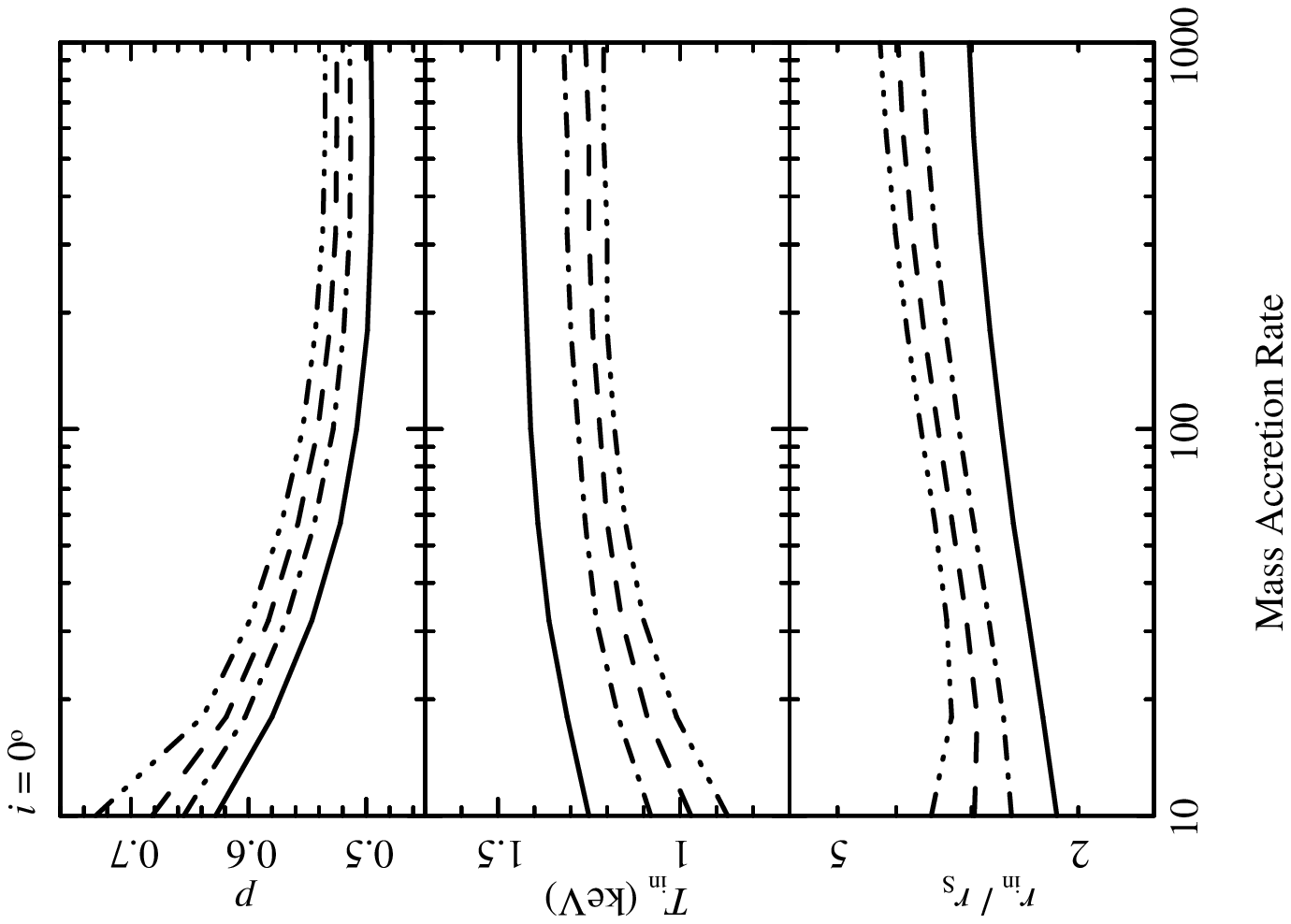}
 \includegraphics[angle=270,scale=0.8]{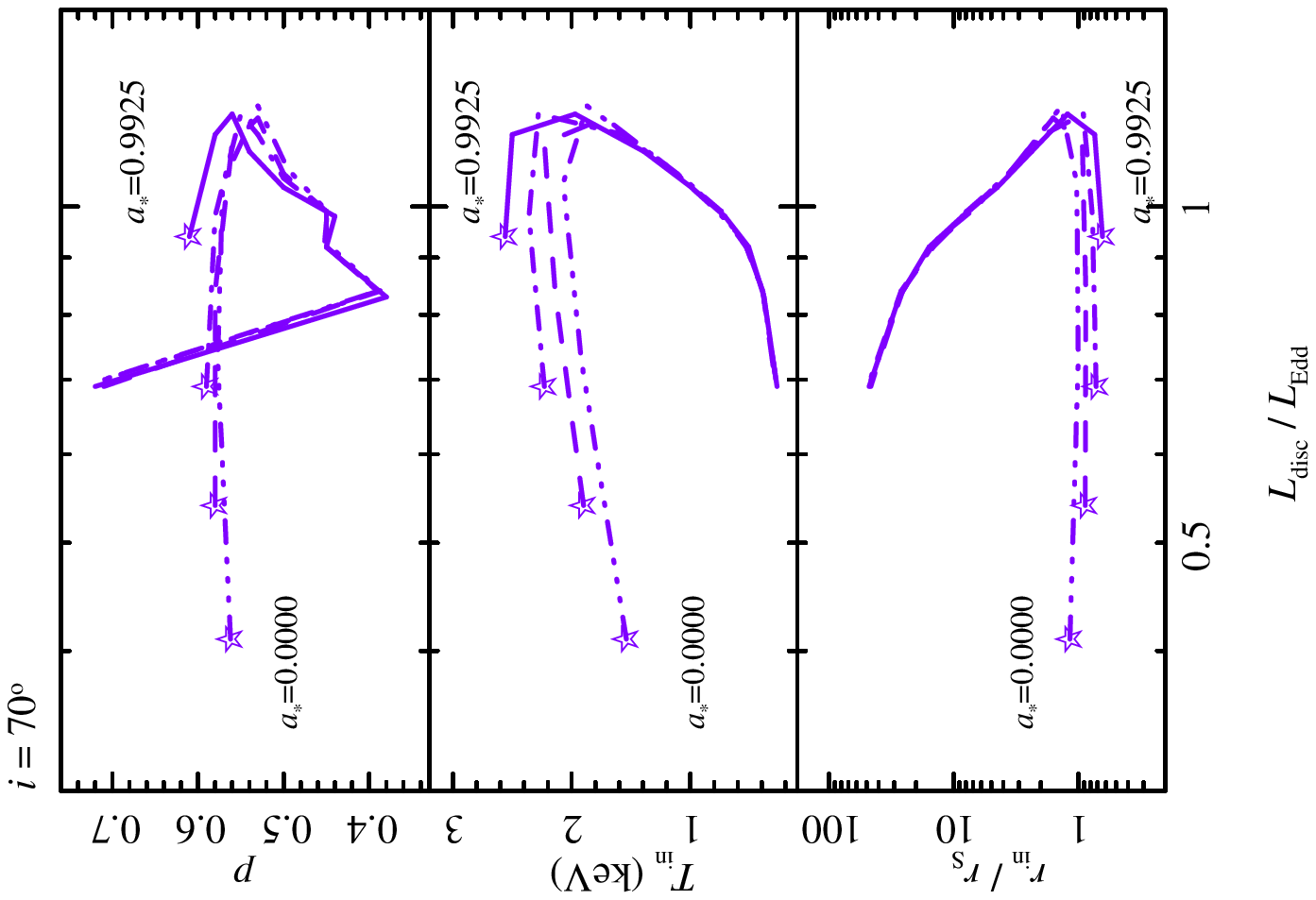}
 \includegraphics[angle=270,scale=0.8]{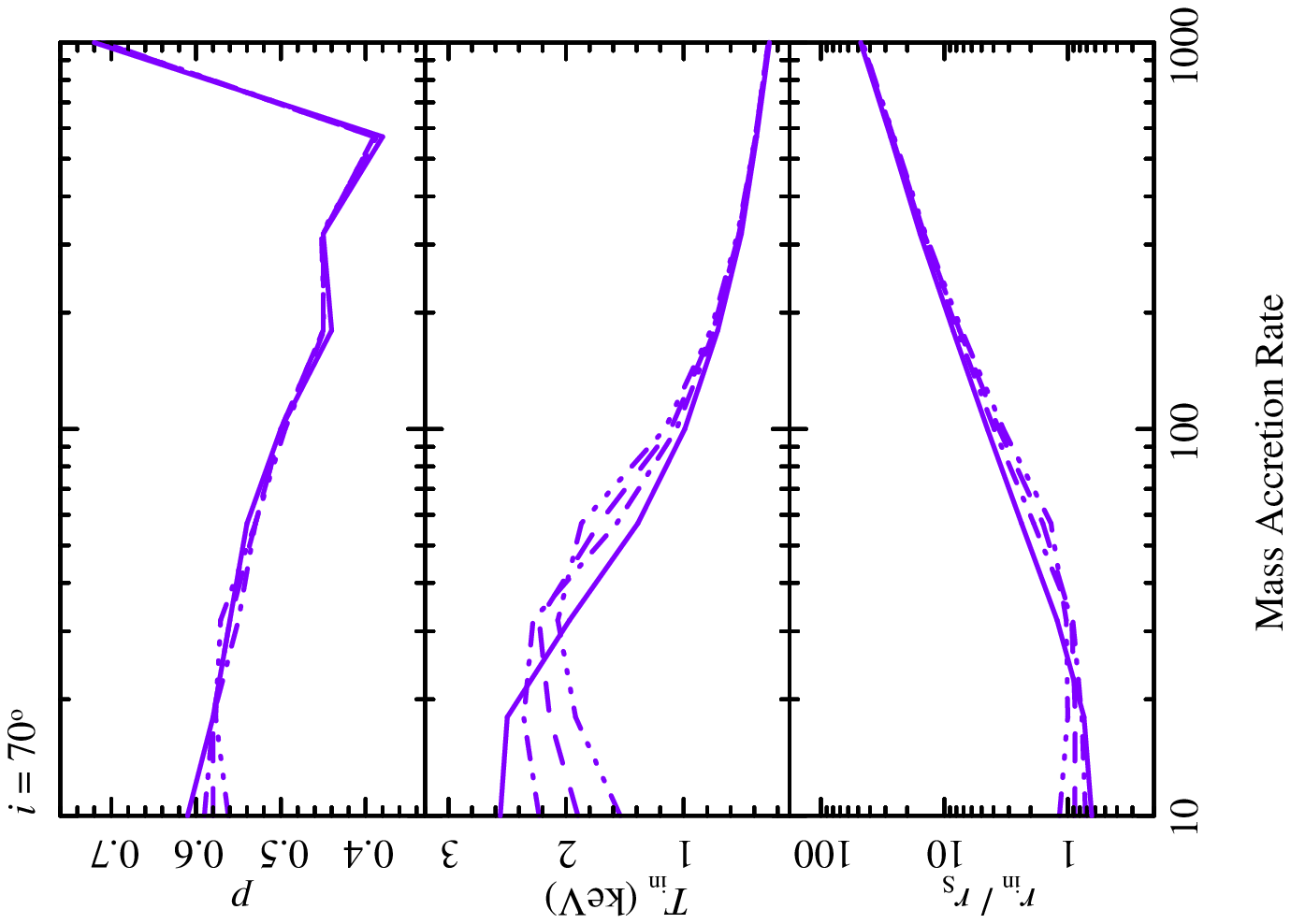}
  \caption{Fitting results for two viewing angles: $i=0$ (upper panels) and $70^{\circ}$ (lower panels), and for various spin parameters: $a_{\ast}=0.0000$, 0.3825,
0.7200 \& 0.9925 in dash-triple-dot, dash, dash-dot, and solid line, respectively.
The stars mark the value when $\dot{m}=10$.
See text for the definition of mass accretion rate ($x$-axis).}
\label{Figure:3}
\end{figure*}

\section{Observable Signatures of the Relativistic Slim Disc}
We fit the synthetic spectra of the relativistic slim disc model with the
extended disc blackbody (extended DBB) fitting model,
the so-called $p$--free model
\citep{27}, at 3 -- 20 keV range.
The parameters of the spectrum model used in our study are the black hole
mass ($m=10\ \rm{M_{\odot}}$), mass accretion rate ($\dot{m}=10,\ 18,\ 32,\ 57,\ 100,\ 180,\
320,\ 570,\ 1000$), dimensionless spin parameter
($a_{\ast}=0.0000,\ 0.3825,\ 0.7200,\ 0.9925$),
and viewing angle ($i=0$ -- $70^{\circ}$).
Hereafter, we will write $a_{\ast}=0$ instead of $a_{\ast}=0.0000$, for the case of non-rotating black hole, throughout the text (except in figures and tables) for the
readers' convenience.

The extended disc blackbody has three fitting parameters: $T_{\rm in}$,
$r_{\rm max}$, and $p$ which are the maximum temperature of the disc, the
radius where $T=T_{\rm in}$, and the temperature gradient, respectively.
In our present study we put most interest in the radius
which defines the size of the emitting region, denoted as $r_{\rm in}$, which is
most useful for comparison with observations.
We calculate $r_{\rm in}$ using disc luminosity and disc effective
temperature, $T_{\rm eff}$.
The disc luminosity is calculated from the theoretical spectrum within
3 -- 20 keV energy range.
The effective temperature is calculated from the maximum disc temperature,
$T_{\rm eff}=T_{\rm in}/ \kappa$, with $\kappa$ being a fixed spectral hardening
factor. We use $\kappa=1.7$, as used in the model calculation. In addition to
$\kappa$, relativistic correction factor $\xi=0.412$
is included when calculating $r_{\rm in}$, implying that the maximum temperature
occurs at somewhat larger radius
\citep{20}.
Thus, the formula to calculate $r_{\rm in}$ can be written as
\begin{equation}
L_{\rm disc}= 4 \pi (r_{\rm in}/\xi)^2 \sigma (T_{\rm in}/\kappa)^4.
\end{equation}
Note, however, that $\xi=0.412$ is derived from analytical Newtonian non-transonic
solution. Lower value of $\xi$ is obtained from pseudo-Newtonian transonic solution
(e.g. Vierdayanti, Watarai \& Mineshige 2008). Therefore, $r_{\rm in}$ in here
should be regarded as an upper limit.
We present our fitting results in Fig. 3 \& 6 and Tables 1 -- 4.

\subsection{The dependence on mass accretion rate}
The dependence of the fitting results on mass accretion rate are shown in Tables 1 -- 4 and
Fig. 3, for two viewing angles, $i=0$
(upper) \& $70^{\circ}$ (lower).
For other viewing angles see Section 3.3.
In Fig. 3, the vertical axes show $p$, $T_{\rm in}$ (keV), and $r_{\rm in}$ ($r_{\rm S}$), in
top, middle and bottom panels, respectively.
The horizontal axis shows disc luminosity and mass accretion rate
in left and right panels, respectively.
Hereafter, the solid, dash-dot, dash, dash-triple-dot lines represent
$a_{\ast}=0.9925,\ 0.7200,\ 0.3825,\ 0$, respectively, unless stated otherwise.
The stars, in the lower left figure, mark the values for the lowest mass
accretion rate, $\dot{m}=10$.

\subsubsection{Face-on case}
We will first discuss the fitting results of face-on case.
For a fixed value of $a_{\ast}$, the effective temperature gradient, $p$, decreases
as the mass accretion rate increases.
That is, the effective temperature profile becomes flatter as $\dot{m}$
increases, as normally expected when the advective cooling becomes important.

As $\dot{m}$ increases, for a fixed value of $a_{\ast}$,
the value of $T_{\rm in}$ increases slowly and tend to saturate when
$\dot{m}$ is greater than 100.
The saturation is caused by energy advection (photon trapping). Due to
energy advection, emergent flux profile in the inner parts is no
longer dependent on $\dot{m}$ (see the upper right part of the spectra
at high $\dot{m}$ in the right panel of Fig. 1).
In addition, the saturation can also be interpreted as an evidence that the disc has reached
the smallest disc radius, $r_{\rm max}$, from which a maximum possible value of flux can be
emitted. This smallest radius is not necessary the inner edge of the disc
whose estimation is not trivial. Also, it may differ from $r_{\rm in}$,
the size of the emitting region.
The radii $r_{\rm max}$ at which $T=T_{\rm in}$ are presented in Table
1 -- 4 (last columns) for different values of $\dot{m}$ and $a_{\ast}$.

We find that the size of the emitting region, $r_{\rm in}$, increases with $\dot{m}$.
Both $T_{\rm in}$ and $L_{\rm disc}$ increase as $\dot{m}$ increases. However, $T_{\rm in}$ increases much slower than $L_{\rm disc}$ which causes $r_{\rm in}$ to increase (since $r_{\rm in}^2 T_{\rm in}^4 \propto
L_{\rm disc}$).
As can be seen in Fig. 3 (upper right figure, middle panel), the
temperature value starts to saturate at $\dot{m} < 100$. However,
Fig. 4 (black lines) shows that the values of $L_{\rm disc}$ continue to increase as $\dot{m}$ increases.
The increase of $L_{\rm disc}$ is not always
proportional to that of $\dot{m}$.
Instead, it grows slower than $\dot{m}$ due to advective cooling.

\begin{table*}
 \begin{center}
  \caption{The fitting results for various $\dot{m}$ for $i=0^{\circ}$,
$a_{\ast}=0.0000$ \& 0.3825.
}\label{tab:first}
     \begin{tabular}{@{}crccccc@{}}
\hline
&&&&&&\\
  $a_{\ast}$ & \ \ $\dot{m}$\ \   & $p$ & $T_{\rm in}$ (keV) & $r_{\rm in}$ ($r_{\rm S}$) & $L_{\rm disc}/L_{\rm Edd}$ & $r_{\rm max}$ ($r_{\rm S}$)\\[2ex]
  \hline
&&&&&&\\
 0.0000  & 10   & 0.730 & 0.87 & 3.50 & 0.37 & 4.64 \\[1ex]
         & 18   & 0.639 & 1.01 & 3.24 & 0.59 & 3.65 \\[1ex]
         & 32   & 0.599 & 1.10 & 3.30 & 0.85 & 3.31 \\[1ex]
         & 57   & 0.572 & 1.15 & 3.45 & 1.12 & 3.14 \\[1ex]
         & 100  & 0.554 & 1.18 & 3.64 & 1.37 & 3.05 \\[1ex]
         & 180  & 0.543 & 1.20 & 3.84 & 1.62 & 3.00 \\[1ex]
         & 320  & 0.537 & 1.20 & 4.01 & 1.83 & 2.94 \\[1ex]
         & 570  & 0.535 & 1.21 & 4.15 & 2.00 & 2.91 \\[1ex]
         & 1000 & 0.535 & 1.21 & 4.25 & 2.11 & 2.91 \\[1ex]
 \hline
&&&&&&\\
 0.3825  & 10   & 0.682 & 0.97 & 2.97 & 0.42 & 3.68 \\[1ex]
         & 18   & 0.619 & 1.09 & 2.94 & 0.65 & 3.17 \\[1ex]
         & 32   & 0.583 & 1.16 & 3.06 & 0.90 & 2.97 \\[1ex]
         & 57   & 0.558 & 1.20 & 3.23 & 1.16 & 2.83 \\[1ex]
         & 100  & 0.541 & 1.22 & 3.41 & 1.41 & 2.75 \\[1ex]
         & 180  & 0.531 & 1.24 & 3.60 & 1.65 & 2.68 \\[1ex]
         & 320  & 0.526 & 1.25 & 3.76 & 1.85 & 2.65 \\[1ex]
         & 570  & 0.525 & 1.25 & 3.89 & 2.01 & 2.63 \\[1ex]
         & 1000 & 0.525 & 1.26 & 3.97 & 2.12 & 2.62 \\[1ex]
 \hline
\end{tabular}
\end{center}
\end{table*}
\begin{table*}
  \caption{The fitting results for various $\dot{m}$ for $i=0^{\circ}$,
$a_{\ast}=0.7200$ \& 0.9925.
}\label{tab:first2}
  \begin{center}
    \begin{tabular}{@{}crccccc@{}}
\hline
&&&&&&\\
  $a_{\ast}$ & \ \ $\dot{m}$\ \  & $p$ & $T_{\rm in}$ (keV) & $r_{\rm in}$ ($r_{\rm S}$) & $L_{\rm disc}/L_{\rm Edd}$ & $r_{\rm max}$ ($r_{\rm S}$)\\[2ex]
  \hline
&&&&&&\\
 0.7200  & 10   & 0.655 & 1.08 & 2.58 & 0.49 & 3.05 \\[1ex]
         & 18   & 0.603 & 1.17 & 2.66 & 0.71 & 2.77 \\[1ex]
         & 32   & 0.568 & 1.23 & 2.81 & 0.96 & 2.62 \\[1ex]
         & 57   & 0.544 & 1.26 & 2.98 & 1.21 & 2.50 \\[1ex]
         & 100  & 0.528 & 1.28 & 3.15 & 1.45 & 2.42 \\[1ex]
         & 180  & 0.519 & 1.30 & 3.31 & 1.68 & 2.35 \\[1ex]
         & 320  & 0.514 & 1.31 & 3.45 & 1.88 & 2.32 \\[1ex]
         & 570  & 0.513 & 1.31 & 3.56 & 2.03 & 2.29 \\[1ex]
         & 1000 & 0.514 & 1.32 & 3.64 & 2.13 & 2.29 \\[1ex]
\hline
&&&&&&\\
 0.9925  & 10   & 0.628 & 1.25 & 2.17 & 0.61 & 2.43 \\[1ex]
         & 18   & 0.580 & 1.31 & 2.29 & 0.84 & 2.25 \\[1ex]
         & 32   & 0.546 & 1.36 & 2.42 & 1.07 & 2.10 \\[1ex]
         & 57   & 0.522 & 1.39 & 2.56 & 1.31 & 1.98 \\[1ex]
         & 100  & 0.508 & 1.41 & 2.68 & 1.53 & 1.89 \\[1ex]
         & 180  & 0.499 & 1.42 & 2.80 & 1.74 & 1.83 \\[1ex]
         & 320  & 0.496 & 1.43 & 2.90 & 1.92 & 1.78 \\[1ex]
         & 570  & 0.495 & 1.44 & 2.98 & 2.07 & 1.76 \\[1ex]
         & 1000 & 0.496 & 1.44 & 3.03 & 2.16 & 1.75 \\[1ex]
 \hline
\end{tabular}
  \end{center}
\end{table*}

\begin{figure}
 \includegraphics[angle=270,scale=0.6]{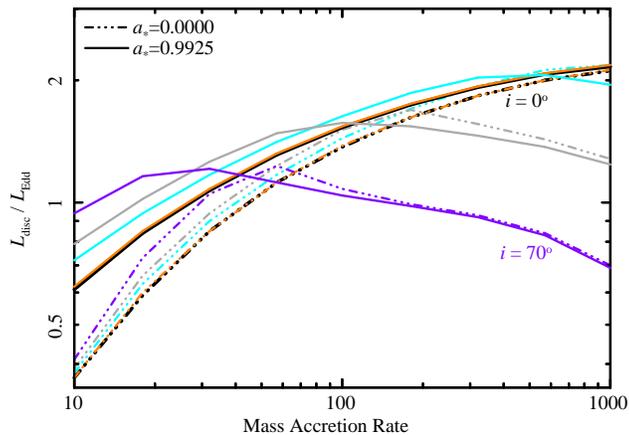}
  \caption{Disc luminosity vs. mass accretion rate for $i=0$, 20, 50,
60 \& $70^{\circ}$ in black, orange, cyan, light gray and purple,
respectively. The energy range for the luminosity calculation
is 3 -- 20 keV.}
\label{Figure:4}
\end{figure}

\begin{figure}
 \includegraphics[angle=270,scale=0.6]{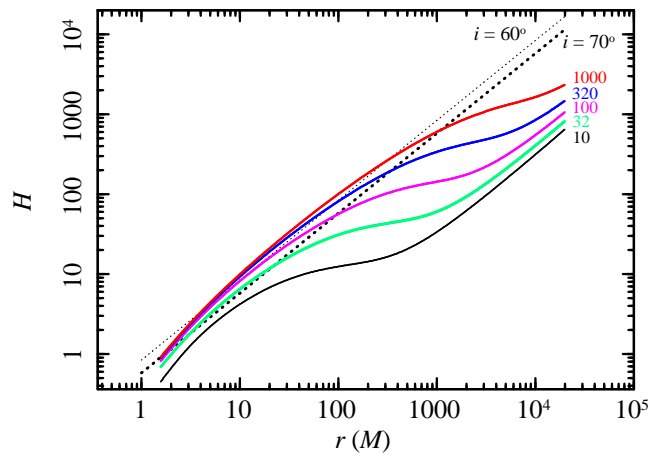}
  \caption{Scale height of the relativistic slim disc model for $a_{\ast}=0.9925$
(solid lines). The colors represent the mass accretion rate (see Fig. 1). The dot lines
represent the observer's direction (viewing angles). The thin line for $i=60$
and the thick line for $i=70^{\circ}$.}
\label{Figure:5}
\end{figure}
\begin{table*}
  \caption{The fitting results for various $\dot{m}$ for $i=70^{\circ}$,
$a_{\ast}=0.0000$ \& 0.3825. The big gap between the rows shows the beginning of the decrese of $L_{\rm disc}$.
}\label{tab:first3}
  \begin{center}
    \begin{tabular}{@{}crccccc@{}}
\hline
&&&&&&\\
  $a_{\ast}$ & \ \ $\dot{m}$\ \  & $p$ & $T_{\rm in}$ (keV) & $r_{\rm in}$ ($r_{\rm S}$) & $L_{\rm disc}/L_{\rm Edd}$ & $r_{\rm max}$ ($r_{\rm S}$)\\[2ex]
  \hline
&&&&&&\\
 0.0000  & 10   & 0.562 & 1.54 & 1.17 & 0.41 & 1.10 \\[1ex]
         & 18   & 0.577 & 1.92 & 1.00 & 0.73 & 0.95 \\[1ex]
         & 32   & 0.571 & 2.07 & 1.03 & 1.05 & 0.95 \\[1ex]
         & 57   & 0.530 & 1.87 & 1.37 & 1.23 & 1.12 \\[3ex]
         & 100  & 0.495 & 1.16 & 3.33 & 1.08 & 2.60 \\[1ex]
         & 180  & 0.449 & 0.77 & 7.24 & 0.99 & 5.27 \\[1ex]
         & 320  & 0.452 & 0.54 & 14.2 & 0.93 & 11.1 \\[1ex]
         & 570  & 0.387 & 0.39 & 25.6 & 0.84 & 18.2 \\[1ex]
         & 1000 & 0.709 & 0.28 & 46.0 & 0.70 & 66.3 \\[1ex]
 \hline
&&&&&&\\
 0.3825  & 10   & 0.580 & 1.90 & 0.88 & 0.54 & 0.86 \\[1ex]
         & 18   & 0.580 & 2.14 & 0.87 & 0.84 & 0.83 \\[1ex]
         & 32   & 0.560 & 2.23 & 0.93 & 1.14 & 0.84 \\[1ex]
         & 57   & 0.530 & 1.72 & 1.59 & 1.20 & 1.32 \\[3ex]
         & 100  & 0.500 & 1.11 & 3.62 & 1.07 & 2.87 \\[1ex]
         & 180  & 0.450 & 0.75 & 7.54 & 0.99 & 5.51 \\[1ex]
         & 320  & 0.450 & 0.53 & 14.6 & 0.92 & 11.5 \\[1ex]
         & 570  & 0.390 & 0.39 & 26.0 & 0.84 & 18.6 \\[1ex]
         & 1000 & 0.710 & 0.28 & 46.3 & 0.69 & 67.2 \\[1ex]
 \hline
\end{tabular}
  \end{center}
\end{table*}
\begin{table*}
  \caption{The fitting results for various $\dot{m}$ for $i=70^{\circ}$,
$a_{\ast}=0.7200$ \& 0.9925. The big gap between the rows shows the beginning of the decrese of $L_{\rm disc}$.
}\label{tab:first4}
  \begin{center}
    \begin{tabular}{@{}crccccc@{}}
\hline
&&&&&&\\
  $a_{\ast}$ & \ \ $\dot{m}$\ \  & $p$ & $T_{\rm in}$ (keV) & $r_{\rm in}$ ($r_{\rm S}$) & $L_{\rm disc}/L_{\rm Edd}$ & $r_{\rm max}$ ($r_{\rm S}$)\\[2ex]
  \hline
&&&&&&\\
 0.7200  & 10   & 0.590 & 2.23 & 0.72 & 0.69 & 0.72 \\[1ex]
         & 18   & 0.580 & 2.36 & 0.77 & 0.98 & 0.74 \\[1ex]
         & 32   & 0.550 & 2.28 & 0.91 & 1.21 & 0.80 \\[3ex]
         & 57   & 0.530 & 1.57 & 1.89 & 1.16 & 1.60 \\[1ex]
         & 100  & 0.500 & 1.06 & 3.95 & 1.06 & 3.17 \\[1ex]
         & 180  & 0.450 & 0.74 & 7.88 & 0.99 & 5.85 \\[1ex]
         & 320  & 0.450 & 0.53 & 15.0 & 0.92 & 11.9 \\[1ex]
         & 570  & 0.390 & 0.39 & 26.5 & 0.84 & 19.0 \\[1ex]
         & 1000 & 0.710 & 0.28 & 46.8 & 0.70 & 67.7 \\[1ex]
\hline
&&&&&&\\
 0.9925  & 10   & 0.610 & 2.56 & 0.64 & 0.94 & 0.67 \\[1ex]
         & 18   & 0.580 & 2.50 & 0.74 & 1.16 & 0.72 \\[1ex]
         & 32   & 0.560 & 1.97 & 1.22 & 1.21 & 1.13 \\[3ex]
         & 57   & 0.540 & 1.39 & 2.37 & 1.12 & 2.09 \\[1ex]
         & 100  & 0.500 & 0.99 & 4.49 & 1.04 & 3.69 \\[1ex]
         & 180  & 0.440 & 0.71 & 8.51 & 0.98 & 6.34 \\[1ex]
         & 320  & 0.450 & 0.51 & 15.7 & 0.92 & 12.6 \\[1ex]
         & 570  & 0.380 & 0.38 & 27.3 & 0.83 & 19.3 \\[1ex]
         & 1000 & 0.720 & 0.27 & 47.6 & 0.69 & 70.1 \\[1ex]
 \hline
\end{tabular}
  \end{center}
\end{table*}

\begin{figure*}
 \includegraphics[angle=270,scale=0.7]{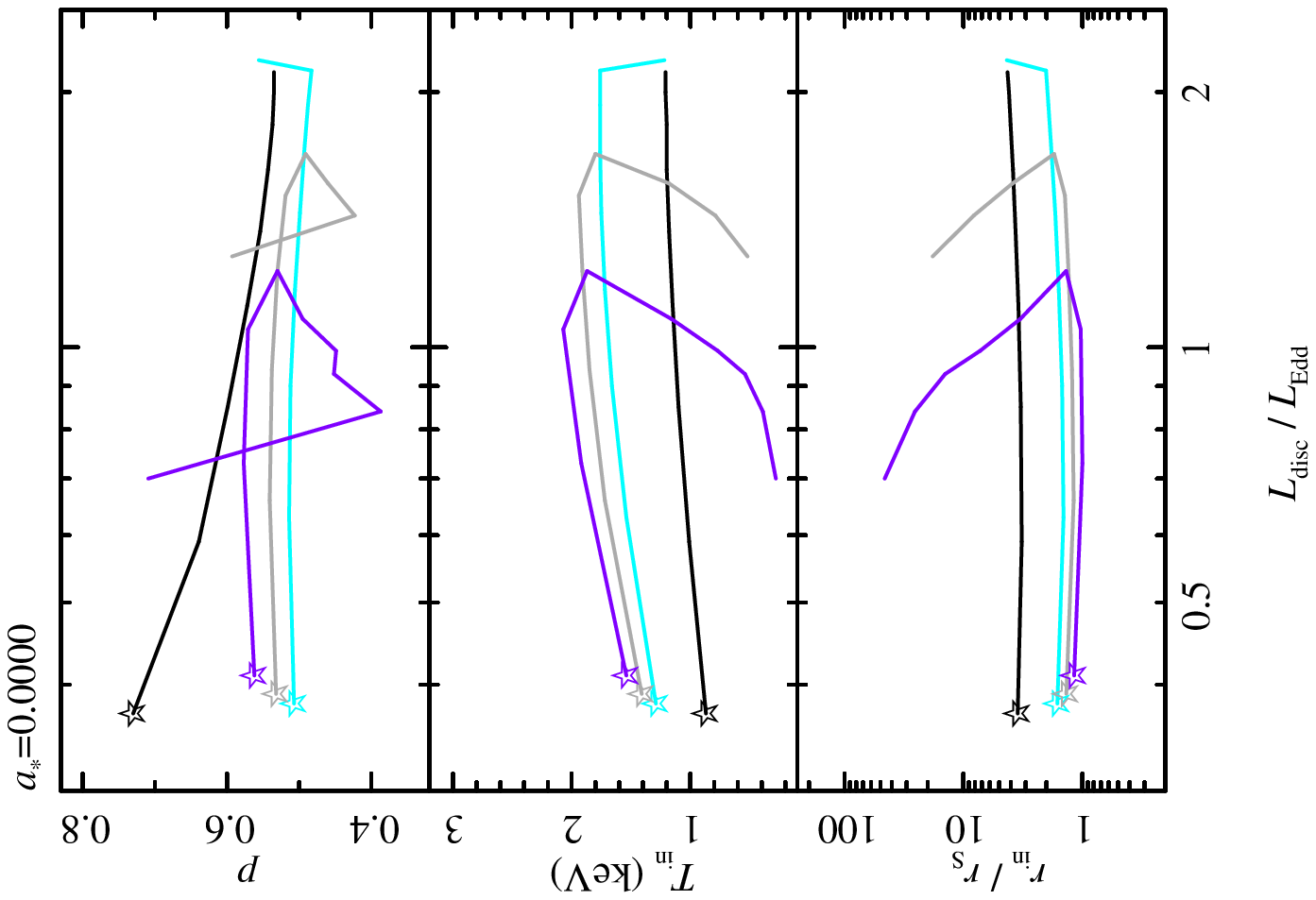}
 \includegraphics[angle=270,scale=0.7]{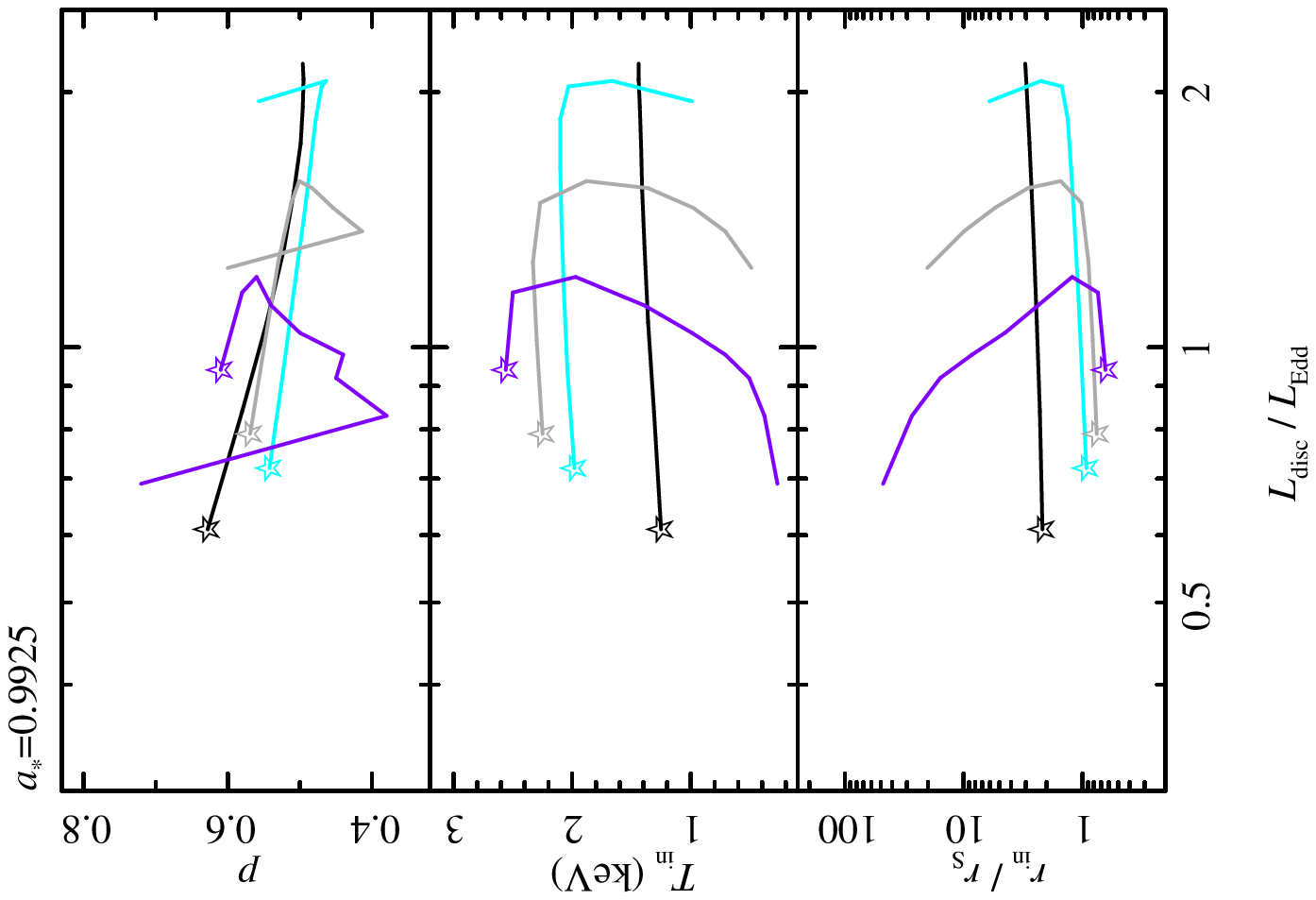}
  \caption{Fitting results for various viewing angles: $i=0$, 50, 60
\& $70^{\circ}$ in black, cyan, light gray and purple, respectively,
for two spin parameters: $a_{\ast}=0.0000$ (left) and 0.9925 (right), both
in solid lines.
The stars mark the value when $\dot{m}=10$.
}\label{Figure:6}
\end{figure*}

\begin{figure*}
 \includegraphics[angle=270,scale=0.6]{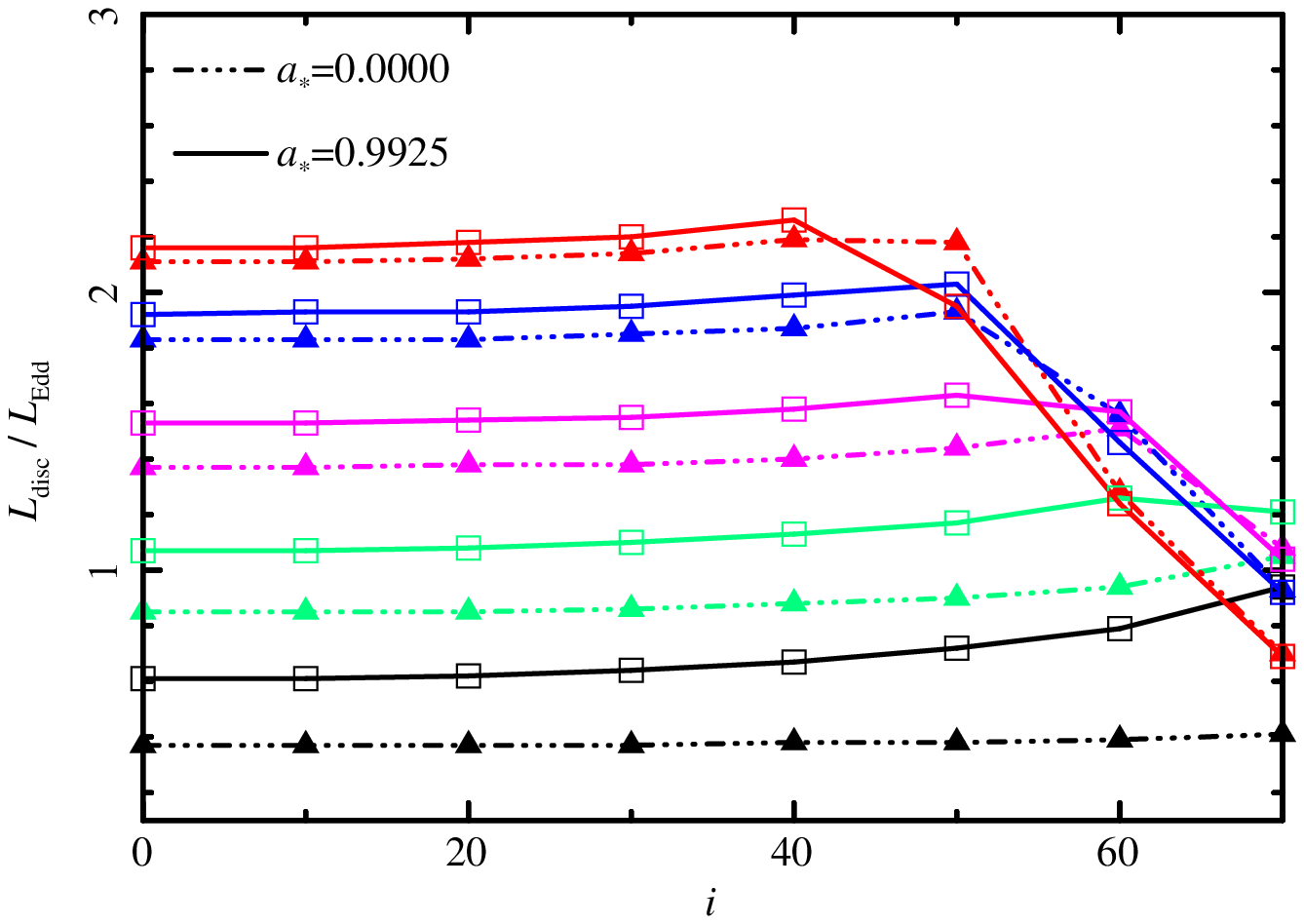}
 \includegraphics[angle=270,scale=0.6]{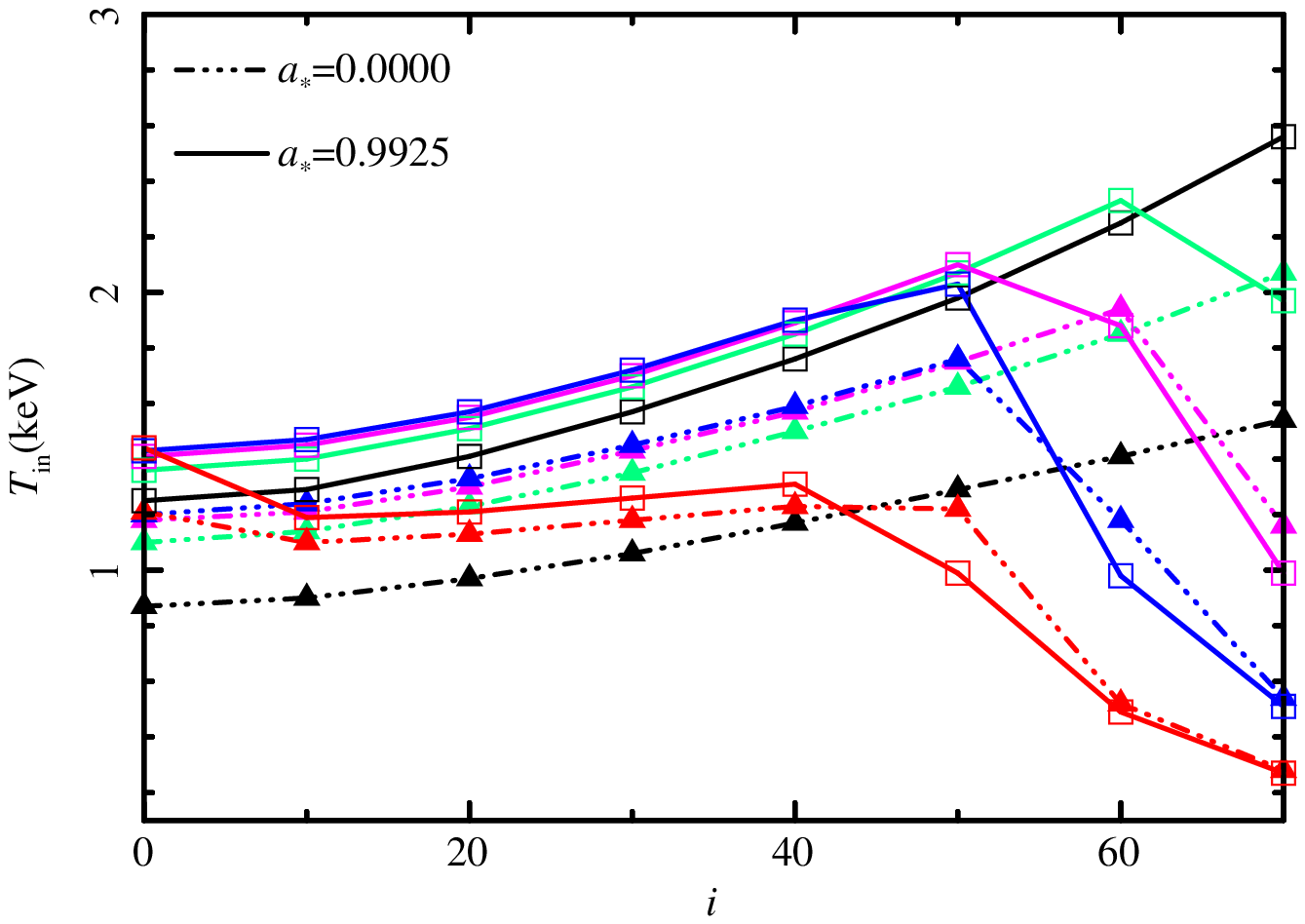}
  \caption{The dependence of the disc luminosity (left) and inner disc
temperature (right) on the viewing angles for various $\dot{m}$: 10 (black),
32 (green), 100 (magenta), 320(blue), 1000 (red).
}\label{Figure:7}
\end{figure*}
\subsubsection{High viewing angle case, $i=70^{\circ}$}
For $i=70^{\circ}$
and at the highest end of $\dot{m}$ we need to set a shorter energy range and
also shift the lower energy range to the softer band to obtain good fits.
For $\dot{m}=1000$, for example, the fitting was only
successful for a very narrow energy range, 0.3 -- 5 keV.
This is caused by the change in the spectral shape as
$\dot{m}$ increases, i.e. becomes softer and the peak
is shifted to softer energy band (see Fig. 2 for
comparison between $\dot{m}=10$ (left) and $\dot{m}=1000$
(right)).
This, in turns, causes a sudden increase in $p$ for $\dot{m}=1000$.
Below $\dot{m}=1000$, $p$-value tends to decrease as $\dot{m}$ increases.

The interesting feature of $i=70^{\circ}$ is the existence of the turning point
at a certain $\dot{m}$ denoted as $\dot{m}_{\rm turn}$.
The effect of the spin parameter and viewing angle on $\dot{m}_{\rm turn}$
will be discussed in the next subsection.
For a fixed value of $a_{\ast}$, $T_{\rm in}$ increases with $\dot{m}$ up to
$\dot{m}_{\rm turn}$, similarly to the $i=0^{\circ}$ case. As $\dot{m}$ increases
further ($\dot{m} > \dot{m}_{\rm turn}$), $T_{\rm in}$ decreases, in contrast to
the face on case (Fig. 3, lower right figure, middle panel).
In the slim disc model, the disc geometry changes, disc becomes thicker as
$\dot{m}$ increases as shown in Fig. 5 where we only plot the
half thickness of the disc for the spin value of 0.9925 (solid lines).
The dotted lines show the observer's line of sight at the viewing angles of
$60$ and $70^{\circ}$.
As the radiation from the innermost hot part of the disc is blocked by the
puffed up radiation-pressure dominated region,
$T_{\rm in}$ decreases significantly.
Similarly to $T_{\rm in}$, $L_{\rm disc}$ also decreases (see Fig. 3 (lower left figure) and Fig. 4 (purple lines)).

Significant increase in the size of the emitting region, $r_{\rm in}$, can be seen
especially for high $\dot{m}$ what can be explained by the obscuration effect.
Due to obscuration, $T_{\rm in}$ decreases much faster than $L_{\rm disc}$,
as shown in Fig. 3 lower (left panel) and Fig. 4 (purple), which
results in the increase of $r_{\rm in}$. Note, however, that $r_{\rm max}$ also
increases when
$\dot{m} > \dot{m}_{\rm turn}$ what shows that we can only observe the
radiation-pressure dominated, cooler part of the disc, surrounding the innermost hot part.

\subsection{The dependence on the spin parameter}
\subsubsection{Face-on case}
For a fixed value of $\dot{m}$, $p$ value becomes smaller for increasing
$a_{\ast}$. When $a_{\ast} \approx 1$, the change in the value of $p$ due to the
change in $\dot{m}$ becomes less significant due to the decreasing value of
ISCO radius as $a_{\ast}$ increases.

The larger $a_{\ast}$ is, the higher becomes $T_{\rm in}$. This fact demonstrates
that there is significant radiation from smaller radii.
In fact, as shown in Table 1 \& 2, $r_{\rm max}$ decreases
as $a_{\ast}$ increases (for a fixed value of $\dot{m}$).
Note, however, that the relativistic effects play a significant role as
$a_{\ast}$ increases.
That is, $T_{\rm in}$ may not be the real maximum temperature of the disc
(see Discussion).

The maximum values of $T_{\rm in}$ in the non-rotating case are around
$T_{\rm in}=1.21$ keV, lower than those of W00--01, even when $\dot{m}=1000$
(see Table 1 \& 2).
Therefore, in order to explain high disc temperature, $\sim 1.5$ keV or
higher, as observed in GRS 1915+105, we will need a
rotating black hole and a high viewing angle (see Table 3 \& 4).
In the present
study, we only consider near-Eddington case that can be regarded as an
extension of McClintock et al. (2006) in which only $L < 0.3L_{\rm Edd}$ cases
are considered.
The size of the emitting region is smaller as $a_{\ast}$ increases which
shows the significant but subtle effect of the light bending.
For a fixed value of $\dot{m}$, the disc luminosity increases but only
slightly,
as $a_{\ast}$ increases.
The observer will receive fewer number of photons when the light bending effect
becomes more important as the inner edge of the disc gets closer to the black
hole.
The innermost disc temperature, $T_{\rm in}$, also increases with
$a_{\ast}$.
The combined dependence of $L_{\rm disc}$ and $T_{\rm in}$ ($L_{\rm disc} \propto
r_{\rm in}^{2} T_{\rm in}^{4}$) on spin result in the decrease of $r_{\rm in}$ as $a_{\ast}$ increases.

\subsubsection{High viewing angle case, $i=70^{\circ}$}
At high viewing angle, the observational features of the relativistic slim
disc spectra become more complicated.
The relativistic effects together with the geometry of the disc play an
important role in determining the observable parameters.
As shown in Fig. 3 (lower left figure), we can see some turning points,
which occur due to the obscuration of the inner, brighter and higher-
temperature part of the disc by radiation-pressure
dominated, cooler part.
It turns out that the turning points appear at different values of $\dot{m}$,
$\dot{m}_{\rm turn}$, whose value depends on the spin parameter.
The higher the spin value is, the lower is $\dot{m}_{\rm turn}$ (see Table 5).
The stars, in the left panel, mark the values at the lowest mass
accretion rate, $\dot{m}=10$.

The value of $p$ is more or less constant, around $\sim 0.5$ -- 0.6, below
the turning point. After the turning points, $p$ decreases, except for
$\dot{m}=1000$ (see 3.1.2).
The trend in the $p$ value may be used to indicate a significant change in
the spectral shape.
However, $p$ seems less sensitive to the spin parameter.

Unlike $p$, $T_{\rm in}$ is rather sensitive to the spin parameter before the
turning points, in which $T_{\rm in}$ is higher for greater spin parameter.
In general, $T_{\rm in}$ increases until it reaches
$\dot{m}=\dot{m}_{\rm turn}$, after which it starts to decrease significantly.
At $\dot{m} > 100$, $T_{\rm in}$ shows no dependence on $a_{\ast}$.

\subsection{The dependence on the viewing angle}
Fig. 6 shows the dependence of the fitting results on the viewing angle
for two spin parameters, $a_{\ast}=0$ \& 0.9925, in the left and right
panels, respectively.

Let us first consider the case when $a_{\ast}=0$.
The fitting results for small viewing angles are very similar.
Also, the values only differ slightly in all $\dot{m}$ range.
A deviation in the trend starts to appear when $i=50^{\circ}$ but only
when $\dot{m}$ is very high $\dot{m} \sim 1000$.
In fact, the deviation becomes more significant at $i=60^{\circ}$.
For $i=60$ \& $70^{\circ}$ case, the trends are very similar
and they look very different from those of the smaller $i$ even when
$\dot{m}$ is not very high, $\sim 100$.

Fig. 7 shows the dependence of $L_{\rm disc}/L_{\rm Edd}$ (left panel) and
$T_{\rm in}$ (right panel) on $i$. The colors represent different values of
$\dot{m}$.
For $a_{\ast}=0$ and for a fixed $\dot{m}$, the value of $T_{\rm in}$ at
$i=70^{\circ}$ is greater than that of the smaller $i$, due mostly to the Doppler
effect.

For small viewing angles, $i < 40^{\circ}$, the turning point never appears.
We only start to see the turning point at $i=50^{\circ}$, at which obscuration
starts to become important.
For $i=40^{\circ}$, the turning point only appears when $a_{\ast}=0.9925$
and $\dot{m}=1000$.

For $a_{\ast}=0.9925$, the situation is quite similar to that of the non-rotating
black hole case. However, a significant difference appears in the value of
$\dot{m}_{\rm turn}$.
As mentioned earlier, the value of $\dot{m}_{\rm turn}$ depends on the spin
parameter as can be seen in Fig. 3 \& 6.
In fact, $\dot{m}_{\rm turn}$ does not only depend on $a_{\ast}$
but also on $i$.
The higher the $i$ is, the lower is the value of $\dot{m}_{\rm turn}$.
The values of $\dot{m}_{\rm turn}$ for $i=50$, 60, and $70^{\circ}$ are presented
in Table 5.
The dependence of $\dot{m}_{\rm turn}$ on $i$ is related to the
geometry of the disc. In other words, we can test the relativistic slim disc
theory by $L_{\rm disc}$--$T_{\rm in}$ diagram obtained from observation.
Dependency of $\dot{m}_{\rm turn}$
on $a_{\ast}$ results from the fact that
the geometry of the inner part of the disc differs significantly
between $a_{\ast}=0$ and 0.9925, for a fixed value of $\dot{m}$.

The increase of $T_{\rm in}$ with
$i$ for $\dot{m}<\dot{m}_{\rm turn}$ can be
explained, again, by the relativistic effects.
The Doppler effect increases the average energy of the photons emitted from
the part of the disc which due to the disc rotation moves toward the observer.
The inner part of the disc across the black hole within the observer's line of
sight can still be seen due to the light bending effect.
The Doppler shift causes a
spectral hardening effect in the observed spectrum due to the increase of the
photon energy which in turns increases $T_{\rm in}$ of the disc.
The light bending effect, on the other hand, affects the observed spectrum
by the increase of the photon flux but without the shift in photon energy.

\begin{table}
  \caption{The value of $\dot{m}_{\rm turn}$ and $L_{\rm disc}/L_{\rm Edd}$ for $i=50$, 60,
$70^{\circ}$, and $a_{\ast}=0.0000$ \& 0.9925.
}\label{tab:second}
  \begin{center}
    \begin{tabular}{@{}ccrc@{}}
\hline
&&&\\
   $i$ & $a_{\ast}$ & $\dot{m}_{\rm turn}$ & $L_{\rm disc}/L_{\rm Edd}$ \\[2ex]
\hline
&&&\\
 $50^{\circ}$ & $0.0000$  & 320  & 1.93  \\[2ex]
 $60^{\circ}$ & $0.0000$  & 100  & 1.51  \\[2ex]
 $70^{\circ}$ & $0.0000$  & 57   & 1.23  \\[1ex]
 \hline
&&&\\
 $50^{\circ}$ & $0.9925$  & 100  & 1.63  \\[2ex]
 $60^{\circ}$ & $0.9925$  & 32   & 1.26  \\[2ex]
 $70^{\circ}$ & $0.9925$  & 18   & 1.16  \\[1ex]
 \hline
\end{tabular}
  \end{center}
\end{table}

\section{Comparison with the Observations of GRS 1915+105}
The observational signatures of the relativistic slim disc presented
in our study should be observed in some sources at high luminosities,
$L_{\rm disc} \sim L_{\rm} \ge 0.3 L_{\rm Edd}$, and when the spectrum is
dominated by the disc emission.
GRS 1915+105 is, so far, the best target of our study, because it maintains
its high luminosity, $L \ge 0.2L_{\rm Edd}$ (Done, Wardzi\'{n}ski
\& Gierli\'{n}ski 2004), since its discovery in 1992 (Castro-Tirado et al.
1994), and also because it has a large viewing angle $\sim 70^{\circ}$
\citep{50}.
The black hole mass in GRS 1915+105 and the distance are estimated to
be $10.1 \pm 0.6 \ \rm{M_{\odot}}$ \citep{53} and $12.5 \pm 1.5$ kpc
\citep{50}.
Similar to other known BHBs, however, the spectra are not always
dominated by disc emission.
Below, we compare our present results with the observations of GRS 1915+105,
focusing on the data with disc-component dominated spectra.

\begin{figure*}
 \includegraphics[angle=270,scale=0.9]{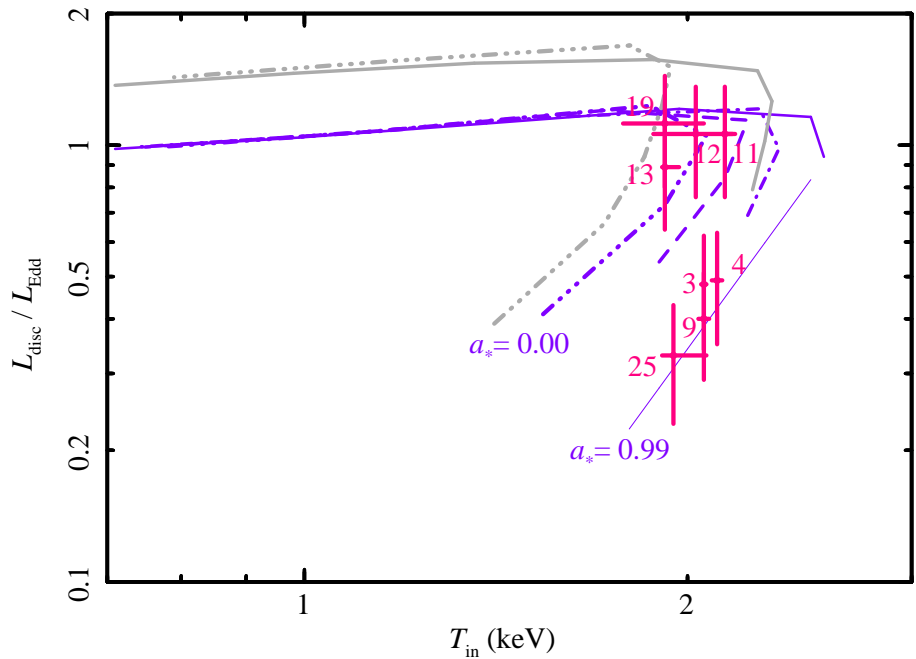}
  \caption{Relativistic slim disc model plotted with spectral fitting
of GRS 1915+105 (Vierdayanti, Mineshige \& Ueda 2010, in magenta).
The color of the lines represent $i=60$ \& $70^{\circ}$ in light gray and purple,
respectively. Thin purple line marks the extension of $a_{\ast}=0.9925$ for $\dot{m}<10$
assuming $L_{\rm disc} \propto T_{\rm in}^{4}$.
}\label{Figure:8}
\end{figure*}

\subsection{GRS 1915+105 at near-Eddington luminosity}
GRS 1915+105 has three unique spectral states, whose transitions
may explain the unique 12 classes of X-ray variability in its light curves
as shown by
\citet{4}
(see also Klein-Wolt et al. 2002; Hannikainen
et al. 2003, 2005). Amongst these classes, there are three quasi-steady
classes whose variability is rather similar to other BHBs, i.e. class
$\chi$, $\delta$ and $\phi$.

Vierdayanti, Mineshige \& Ueda (2010), hereafter VMU2010, studied the behavior
of GRS 1915+105 at near-Eddington luminosity.
They chose data from {\it RXTE} observations taken in 1999 -- 2000.
In order to keep the results applicable for other BHB systems, they only focused
on the data that show little variability.
There were two groups of data in their final sample:
those whose spectrum is dominated by the disc component and those
dominated by the component other than the disc.

They fitted both groups using a disc+corona model. They chose the disc
blackbody (DBB) model and, in addition, they also tried the extended DBB model for the disc-
dominated data.
As for the coronal component, they chose the thermal Comptonization model in which
the reflection of the up-scattered photons by the disc, a 7.0 keV (K$\alpha$)
absorption line of iron and its corresponding edge are taken into account as done
in Ueda, Yamaoka \& Remillard (2009).

VMU2010 found a new branch in the $L_{\rm disc}$ -- $T_{\rm in}$ plane
(X-ray HR diagram,
hereafter) as shown in fig. 11 of VMU2010.
For disc-dominated spectra, they found that there seems to be two branches of
$L_{\rm disc} \propto T_{\rm in}^4$ in the X-ray HR diagram when the disc
blackbody model is applied.
The upper branch has never been observed in other BHBs.
Normally, when the spectra are dominated by the disc component, they
follow the $L_{\rm disc} \propto T_{\rm in}^4$ relation that implies a constant
$r_{\rm in}$.
In the standard disc model scenario, it means that the disc inner edge has
reached the ISCO.

When the extended DBB model is applied, however,
they found that these spectra show deviation from that of the standard
accretion disc, despite being disc-dominated.
The $p$ values were closer to the value of the slim disc model, i.e.,
$p=0.5$ instead of that of the standard disc ($p=0.75$).
Moreover, the higher luminosity branch in their X-ray HR diagram does not
follow $L_{\rm disc} \propto T_{\rm in}^4$.
Instead, it resembled the relativistic slim disc branch at moderately
high $\dot{m}$ and high $i$ as we find in our present study.

\subsection{Interpretation based on relativistic slim disc model}
We plot VMU2010 data
together with the results of this work
 for $i=60$ (light gray) \& $70^{\circ}$ (purple) in Fig. 8.
Regarding the VMU2010 data, we took all the disc-dominated
datasets which satisfy $L_{\rm disc}>L_{\rm Compt}$ when fitted with the extended DBB model, where $L_{\rm Compt}$ is the luminosity of the Comptonization component. These are dataset 3, 4, 9, 11, 12, 13, 19 and 25.
The thin purple line marks the extension of $a_{\ast}=0.9925$ for $\dot{m}<10$
assuming $L_{\rm disc} \propto T_{\rm in}^{4}$.
The most significant result is that we can now explain the origin of the two groups
of data on the $L_{\rm disc}$ -- $T_{\rm in}$ diagram which were discovered by
VMU2010. The lower branch is an extension of the standard disc branch in the
high luminosity regime, while the upper branch is produced by the inner disc
obscuration.

Based on the lower luminosity data points (dataset 3, 4, 9, and 25),
high spin parameter is strongly favored.
As a reminder, solid lines in Fig. 8 represent
$a_{\ast}=0.9925$.
Assuming that GRS 1915+105 has a rapidly spinning black hole, $a_{\ast}=0.9925$,
the turning point due to obscuration occurs at $\dot{m}_{\rm turn}=18$ (see Table 5).
Therefore, the mass accretion rate, $\dot{m}$, for the lower luminosity data points
should be lower than 18. For the upper luminosity data points, $\dot{m}$
should be higher than 18 but lower than 100, since otherwise high value of
$T_{\rm in}$ cannot be explained (see $T_{\rm in}$ in Table 4 for $a_{\ast}=0.9925$
and $\dot{m}>100$).

From our present analysis for $i=70^{\circ}$
and $a_{\ast}=0.9925$ we obtain $p=0.61$ for $\dot{m}=10$ and $p$ becomes
smaller as $\dot{m}$ increases ($p=0.5$ for $\dot{m}=100$).
Note that for lower $a_{\ast}$, $p$ value is already below 0.6 even for $\dot{m}=10$
despite the fact that the range is within the error estimated from
observations.
VMU2010 estimated the $p$ values of the data in which the disc fraction is
significantly high ($> 70 \%$) to be $\sim 0.6$. That is, $p$ values obtained
by VMU2010 are consistent with the case of $\dot{m}=10$ -- 32 in our present
study.

The high maximum disc temperature commonly found in GRS 1915+105 is also an
important signature of slim disc at high viewing angle.
As discussed in Section 3, the observer at higher viewing angle will have
more chance to observe both the higher energy photons as well as the emission from
the smaller radii, when $\dot{m} < \dot{m}_{\rm turn}$.
In the case when $i=70^{\circ}$, the maximum disc temperature can reach 2 keV
or even slightly higher.
We suggest that the disc-dominated data of GRS 1915+105 used in VMU2010
are the first evidence of relativistic slim disc theory and its obscuration at
high spin parameter and viewing angle.

\section{Discussion}
\subsection{Interpretation of $T_{\rm in}$}
The relativistic effects should play a significant role as
the inner edge of the disc gets closer to the black hole.
The estimate of the inner edge radius would require a full knowledge of the gravitational
redshift and the light bending.
It is very likely that this inner edge is
a region that is visually hidden to the observer, from which the emission cannot escape along his line of sight.
It implies that the observed innermost disc temperature, $T_{\rm in}$, may not
be the real maximum temperature of the disc when relativistic effects become
important.

As shown in Table 1 \& 2, $r_{\rm max}$ values saturate at around $\sim 2.9r_{\rm S}$
and $1.75r_{\rm S}$, for $a_{\ast}=0$ and 0.9925, respectively, for the face-on case
even when $\dot{m} > 320$.
From the numerical simulation we know that the inner edge of the disc can reach
$\sim 1.4r_{\rm S}$ for $a_{\ast}=0$ and
$\sim 0.8r_{\rm S}$ for $a_{\ast}=0.9925$ at $\dot{m}=1000$.
Thus, the saturation of $r_{\rm max}$ shows that
the emission from the innermost part at $r<r_{\rm max}$ is not
captured by the extended DBB model even when $i=0^{\circ}$.
Since the innermost part of the disc is invisible, we automatically lose
the information of the real maximum temperature of the disc.
In the case of high-$i$, the situation becomes even more complicated especially at
high $\dot{m}$ due to the obscuration effect.
Unless we know $a_{\ast}$, $\dot{m}$ and $i$ of the observed systems, it is difficult to
estimate the real maximum temperature of the disc from observation.
Ironically, we usually estimate $\dot{m}$ and $a_{\ast}$ from observable parameters such as
$L_{\rm disc}$ and $T_{\rm in}$.

\subsection{Disc Geometry}
So far the comparison between the theory and observations based on spectral fitting
has been made mainly to check the surface temperature profile of the disc.
The present study provides another way of comparison, since it can test not only
the surface temperature profile but also the disc geometry (i.e., the disc height
distribution, $H(r)$).
The observed behavior of the disc on the $L_{\rm disc}$
-- $T_{\rm in}$ diagram can provide an independent support to the relativistic slim disc model.

In the standard picture of accretion disc, the disc is assumed to be
geometrically thin. That is, the disc obscuration becomes less important
except when the viewing angle is $\sim 90^{\circ}$.
In the slim disc model, however, the disc geometry becomes thicker from
that of the standard case when the radiation pressure becomes important.
On the other hand, the photon trapping due to advective motion will
reduce the disc geometrical thickness.
Therefore, as the photon trapping becomes effective, we expect the decrease
of the disc geometrical thickness, as can be seen in Fig. 5.

In Fig. 5, we can also see that the photon trapping starts to
become effective at larger radii as $\dot{m}$ increases.
However, we can also see that the scale height of the disc greatly increases
with $\dot{m}$. It explains why at moderately high viewing angle, we could
never see the inner part of the disc when mass accretion rate is extremely
high, $\dot{m} \sim 1000$.

\subsection{Black Hole Spin Estimate}
Estimating the spin parameter is not an easy task especially if we only
have a short range of luminosity data.
In the case of low viewing angle, the value of the maximum temperature and
disc luminosity may help in distinguishing non-rotating black holes from the
rapidly rotating ones.
In the case of high viewing angle, in addition to $T_{\rm in}$ and $L_{\rm disc}$,
$\dot{m}_{\rm turn}$ is useful in estimating $a_{\ast}$ as $\dot{m}_{\rm turn}$ depends
on $a_{\ast}$ and $i$.
Estimating $\dot{m}$ itself, although possible, is not trivial.
By comparison with the observations, we suggest a spin parameter value,
$a_{\ast} > 0.9$ for GRS 1915+105.
Focusing on low luminosity data, $L_{\rm disc} < 0.3L_{\rm E}$,
\citet{24}
also found $a_{\ast} > 0.98$ for GRS 1915+105.
It is interesting to note in this respect that \citet{26}, due to their selection
criteria, mainly used high luminosity data of GRS 1915+105 and derived lower spin
parameter value, $a_{\ast} \sim 0.7$. We can now understand the reason for this;
if we fit the upper branch data (at high luminosities) by the standard disc theory,
we will get lower spin parameters (see Fig. 8).

We also need to address the absorption (both continuum and line) model issue.
Accurate estimation of the column densities of major elements in the
interstellar matter between the source and the observer becomes
very crucial, especially for GRS 1915+105. It will significantly affect the
energy range of the X-ray emission from BHBs. VMU2010 followed
\citet{39}
in modeling the absorption by the interstellar and
circumstellar gas.
\citet{39} in their study used the data from
simultaneous {\it Chandra} HETGS and {\it RXTE} observations which give
better statistics than some previous studies (e.g. Lee et al. 2002).
The fitting results, however, seem to have a strong dependence on the
absorption model. The results may change when different treatment for the
absorption, i.e. different absorption model, is applied which remains an open
question.

\section{Conclusions}
We have studied the observational signatures of the relativistic slim disc
(S11) by fitting the synthetic spectra with the extended DBB model.
The dependence of these signatures on the accretion rate ($\dot{m}$), black
hole spin ($a_{\ast}$), and inclination angle ($i$) have been
investigated. We have shown that it is possible to estimate those physical
parameters from observations.
We found a new parameter, $\dot{m}_{\rm turn}$, which depends on the
spin parameter for high viewing angles.
We also found that the two branches found by VMU2010 on $L_{\rm disc}$ --
$T_{\rm in}$ plane can be explained by the relativistic slim disc model.
We suggest that the disc-dominated data of GRS 1915+105 used in VMU2010
is the first evidence of the relativistic slim disc and its obscuration
at high spin parameter and viewing angle.

\section*{Acknowledgments}
We gratefully thanks the referees for their comments that help
improved this work. This work is supported in part by research grant
{\lq Program Hibah Bersaing\rq} (DIPA ITB 2011
No.003/TL-J/DIPA/SPK/2011, KV), the Grant-in-Aid of Ministry of Education, Culture, Sports, Science, and Technology (MEXT) (22340045, SM) and by the Grant-in-Aid for the global COE programs on The Next Generation of Physics, Spun from Diversity and Emergence from MEXT (KV, SM). AS was supported in part by NASA grant NNX11AE16G. MB was supported in part by RVO:67985815. KV thanks Jun Toshikawa and Mahasena Putra for fruitful discussion.

\end{document}